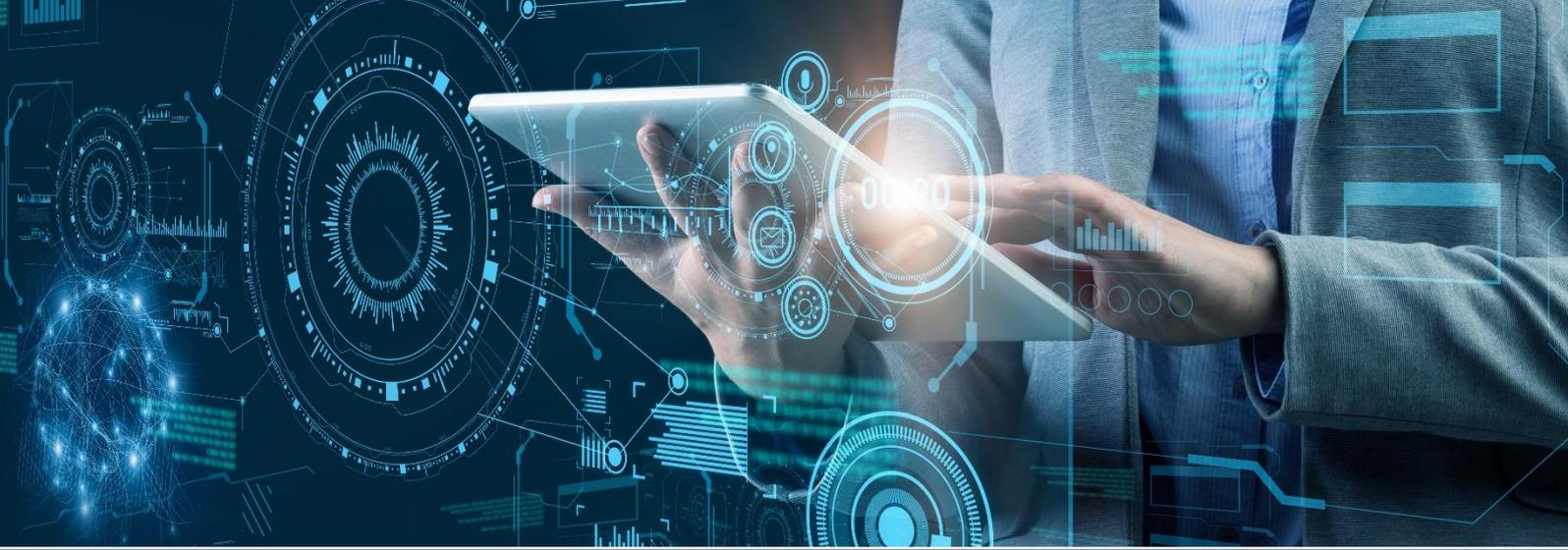

**August 2022**

*Technical Report*

# Socio-Technical Security Modelling:
## Analysis of State-of-the-Art, Application, and Maturity in Critical Industrial Infrastructure Environments/Domains.

**Uchenna D Ani**
**Jeremy Watson CBE**
**Nilufer Tuptuk**
**Steve Hailes**
**Aslam Jawar**

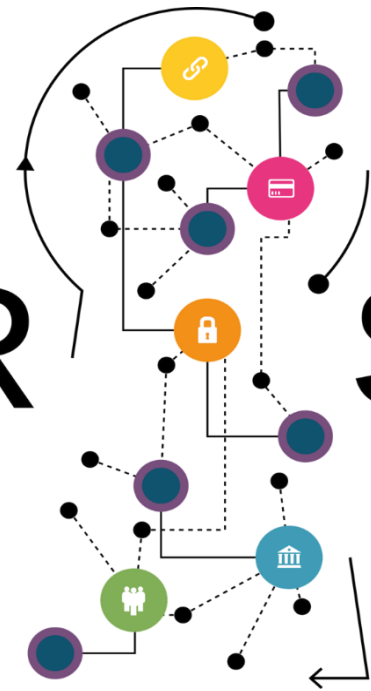

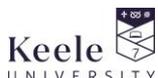
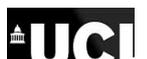

**Modelling for Socio-Technical Security Project Report**
PETRAS National Centre of Excellence in IoT Systems Cybersecurity.

**Details of document preparation and issue:**

| Version no. | Prepared | Checked | Reviewed | Approved | Issue date | Issue status |
|---|---|---|---|---|---|---|
| 1.1 | Uchenna D Ani | | - | - | 20/02/21 | 1$^{st}$ Draft |
| 1.2 | Uchenna D Ani | Nilufer Tuptuk | Jeremy Watson | Jeremy Watson | 20/08/22 | 2$^{nd}$ Draft |
| 1.3 | Uchenna D Ani | | Jeremy Watson | Jeremy Watson | 02/09/22 | 3$^{RD}$ Draft |
| | | | | | | |

# Table of Contents

# List of Figures

# List of Tables

# Executive Summary

The security of cyber-physical system (CPS) infrastructure is proving more complex with the growing digitalisation of organisational processes that see the integration of new computing technologies and trends such as Inter-networking, Internet of Things (IoT), Big Data, Artificial Intelligence, Cloud and Edge Computing, etc., into legacy industrial critical infrastructure systems. These have now enabled the linking of disparate and formally isolated infrastructure systems into a *'System of Systems'* coupled infrastructure. While the digitalisation trends add efficiency benefits to the functions and operations of cyber-physical CIs, it also introduces new security issues. Thus, cyber-enabled attacks targeting cyber-physical CI systems are increasing and are dealing severe harmful impacts to the operations of the systems and the societal functions they support. How to effectively secure or manage the growing security risks continues to be at the fore of engagement and discussions in the critical infrastructure operations and security communities. Although, several security solutions exist for addressing these cyber threats, these appear to skew more towards technologies, often overlooking the non-technical attributes of CPS infrastructures such as human factors, organisational structure, policy, and procedures.

As a modern industrial system reflects an open socio-technical system comprising; goals and values, managerial, psychological, structural, and technical elements that all co-interact within a wider singular environment, successful cyber incidents and emerging threats have shown that technology alone is neither the cause nor the solution to the cybersecurity problems in CPS. Better security can be realised through considerations of how the technical dimensions overlap with the social, and how the growing overlap can influence system security and resilience or otherwise. But accomplishing this on live CI system comes with some operational safety and security challenges and consequences as security audits on such live CI system often cause disruptions to process functions. Modelling and Simulation (M&S) is an approach adopted as a viable alternative for exploring and addressing security-related objectives in such environments. This enables to support the analysis of security risks and with the determination or synthesis of appropriate ways to manage security risks.

This study explores the state-of-the-art, application, and maturity of socio-technical security models for industries and sectors dependent on CI and investigates the gap between academic research and industry practices concerning the modelling of both the social and technical aspects of security. Systematic study and critical analysis of literature show that a steady and growing on socio-technical security M&S approaches is emerging, possibly prompted by the growing recognition that digital systems and workplaces do not only comprise technologies, but also social (human) and sometimes physical elements.

Most of the model representation formats analysed appear as methodologies. This highlights that using underlying theories to define and understand how socio-technical security is being applied to M&S and tends towards achieving usable tools from the methodologies. Information Technology and Telecommunications, Financial Services, Government facilities, Health and Public Health, and Transportation appear to be the top application domains/sectors that are more actively gaining push in exploring the socio-technical dimension of security. Presumably motivated by the view that existing security measures are not sufficiently effective or holistic to counteract or reduce cybersecurity incidents. The existing modest activity in this area by UK critical infrastructure sectors calls for attention and needs response to prevent grave consequences. This means that socio-technical security M&S approaches and tools need to be developed and progressed to maturity. However, such maturity comes from several acknowledgements, and endorsements.

Our study shows that most existing works around socio-technical security M&S for critical infrastructure systems are still in theoretical conceptions and analysis stages, thus not matured enough to be applied to real/live environments. This highlights the endeavours to get socio-technical security modelling right at the idea development and testing stages, and also to allow ample scrutiny and validation before security M&S tools are moved to the real environment.

Also, contrary to expectations, a very little involvement and representation is seen of social science domain experts. This is perhaps due to their lack of engagement with such methods in their typical research and/or their weak understanding of technical security perspectives. For the work reviewed, both technical and social factors and attributes are considered in the design and development of security for infrastructure systems. These attributes appear along a range of categorisations, including technologies, actors or agents, relationships and interactions, policies, cultures and values, resources, or assets available, and organisational structures, each of which can be further broken down into related attributes.

Clearly, the findings highlight a strong link amongst social and technical (and physical) security factors, attributes, and associated sub-attributes that are appropriate for modelling and simulating security within critical infrastructure systems and/or organisations. Often, emerging security conditions involve a combination of one or more technical (and physical) and/or social system attributes, which interact in a system-like approach, and influence both security, functionality, and performance of the system involved. However, there appears to be weaker considerations of the socially-inclined security attributes.

Most of the works exploring socio-technical security appear to be more concerned with the security issues/risks emerging from actors/agents – humans – traits. Emphasis on humans – presumably being the weakest link and/or being the increasing target of cyber threats – appear to be high. For security aspects covered, one area of dominant focus found is on security requirement analysis/engineering for systems or organisations where one or more of human actors/agents, technologies, and organisational elements are involved. Another is the modelling and simulation of interactions between system technology components relating to the tactics for successfully reaching and implementing attacks. A third involves the modelling and simulation of threats to the security and privacy of communications data or information. Three approaches – 'CySeMoL', 'Tropos', and 'SePTA' – appear to stand out in terms of interdependency modelling considerations.

The growing spread of the problems requires efforts towards reducing the requirements for security expertise in the use of security M&S tools while allowing for security evaluations. Also, there is a need to adopt an evolutionary approach to socio-technical security reasoning, design, and implementation to enable continuous evaluations and joint-optimisation of security and resilience in response to the dynamics of security risks. A useful recommendation along this direction involves improving the communication – awareness and briefing activities – amongst stakeholders about socio-technical security issues, risks, and effective countermeasures along socio-technical lines. This would support wider awareness of emerging security risks and timely and constructive decision-making on effective solutions. Also, better security can come from understanding the functions and interactions amongst CNI system functions and interactions relative to implementable security mechanisms. Security features need to be shaped to enable easy translation into active security mechanisms to help advance security-by-design. Government and policy makers can also support the wider awareness and adoption of socio-technical security by developing policies that can shape and signpost the CNI cybersecurity environment. This can persuade compliance, increase confidence and trust in the holistic nature of the system along security viewpoint.

# 1. Introduction

## 1.1 Background – The Issue/Problem

The growing numbers and evolving nature of cyber-attacks [1] have kept issues open on how to effectively secure or manage cyber security risks for modern cyber-physical system (CPS) infrastructure. Cybersecurity risks in CPS are growing and are at the forefront of concerns for the critical infrastructure community. The task of enforcing or ensuring cybersecurity is proving overtly more complicated with new trends that see increased digitalisation of organisational processes [2] through the incorporation of new technologies including Inter-networking, Internet of Things (IoT), Artificial Intelligence, and Edge Computing into traditional industrial critical infrastructure systems [3]. These have transformed older, legacy, and formerly isolated infrastructure systems into a system of coupled complex infrastructure referred to as *'System of Systems'* (SoS) [4]. It is not far-fetched that as more complex, large, and connected systems and environments continue to emerge from these conscious developments, the scale of connectivity and complexity only grows further. This increases the scale of security risks; widening the attack surfaces and increasing attack incidents and potential impacts [5]. By security we mean cybersecurity or the security of internetworked components and systems.

Thus, security risks need to be managed in the face of new and emerging forms of attack vectors and increased sophistication where highly resourced (and motivated) adversaries are capable of disrupting critical societal services to the detriment of a significant proportion of the population [5]. While this may be true, there are growing concerns about how to effectively manage cybersecurity risks when cyber incidents appear to be growing in number, despite the considerable control efforts and solutions available. The misalignment of security management strategies stemming from the lack of a holistic system view appears to be one of the drivers of this phenomenon [6]. This highlights a situation where the non-technical elements or constituents of the cyber-physical system such as humans and organisational structures are often less or not considered as functional parts of the broader system which can make or break security [7], [8]. Often, when reasoning about, and implementing practical security solutions, the attributes of these non-technical system elements are not well considered.

Conversely, the successes of most recent cyber incidents including those targeting industrial control and cyber-physical systems critical infrastructures appear to be enabled by failure scenarios linked to one or more of these non-technical elements. For example, the actions or inactions of humans in the loop or due to weak or lack of consciously secure organisational processes [9], [10]. Misaligned security strategies and measures may explain why, while the security community has been more focused on developing tools and techniques to keep the technical aspects/elements of industrial systems robustly secure, intelligent and highly-resourced malicious actors have gradually changed their attack vector directions towards the less-technical or more social elements and attributes of systems [6], see *Figure 1*. Top amongst these appears to be the human element [9], [10].

This view can be seen in recent cyber-incident accounts. Recently, it is becoming more evident that the capability to cause a significant public breach or compromise of industrial systems appears to be facilitated or enabled more by the incompetence [11], and/or negligence [12] of the target organisation than by the competence and skills of the attacker. Of course, the latter also plays a part in enhancing the capability to compromise. However, aside from the smartness of attackers, well-reported cyber incidents such as Stuxnet that targeted the Iranian Nuclear Power Plant, the attack on Saudi Aramco's Industrial Systems, and the IT-focused WannaCry attack that impacted the UK NHS and SingHealth Singapore [13], [14] all highlight fundamental failures in the affected organisations being at the root of the attack successes [11]. The failures typify demonstrations of a lack of security awareness, threat consciousness, and/or competence within the organisational workforce involved. This is as true for industrial and cyber-physical infrastructure systems as it is for information technology or enterprise systems. In fact, the operational technology community seem less cyber-aware than their enterprise system counterparts.

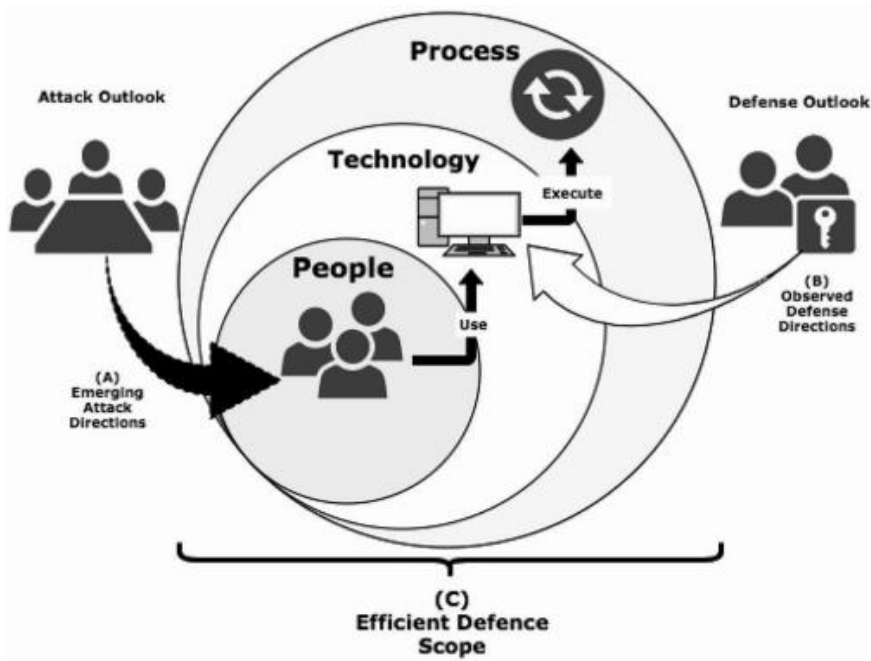

Figure 1: ICS security trends (attack and defence outlook) [6]

## 1.2   A Socio-Technical Security Approach to a Socio-Technical Industrial Organisation

An industrial organisation provides a good example of a socio-technical system since it comprises various subsystems that include social and technical attributes. For example, an industrial organisation can be viewed as an example of an open socio-technical system which comprises five elements; goals and values, managerial, psychological, structural, and technical subsystems, all acting within a wider environment [15] as shown in *Figure 2*. While the first four elements outlined can be mapped to social dimensions or characteristics, the last maps to the technical domain. Each of the five parts possess their own input-conversion-output processes related to, and co-operating with the other subsystems. These interrelationships and interactions hold true for cybersecurity contexts as well as functionality and operational performance aspects.

As shown in *Figure 2*, each of these subsystems maintains certain attributes. For example, Goal and Value attributes can include culture and philosophy. Managerial attributes can include goal setting, planning, and implementation. Psychological attributes can include attitude, perceptions, motivations, behaviours, and communications. Structural attributes can include tasks, procedures, workflow, and rules or policies. Technical attributes can include knowledge, skills, techniques, equipment or tools, and facilities. Each of these subsystem elements and attributes can influence how secure or vulnerable the entire system can be, given that they all interact and contribute to the normal functioning and security of the system.

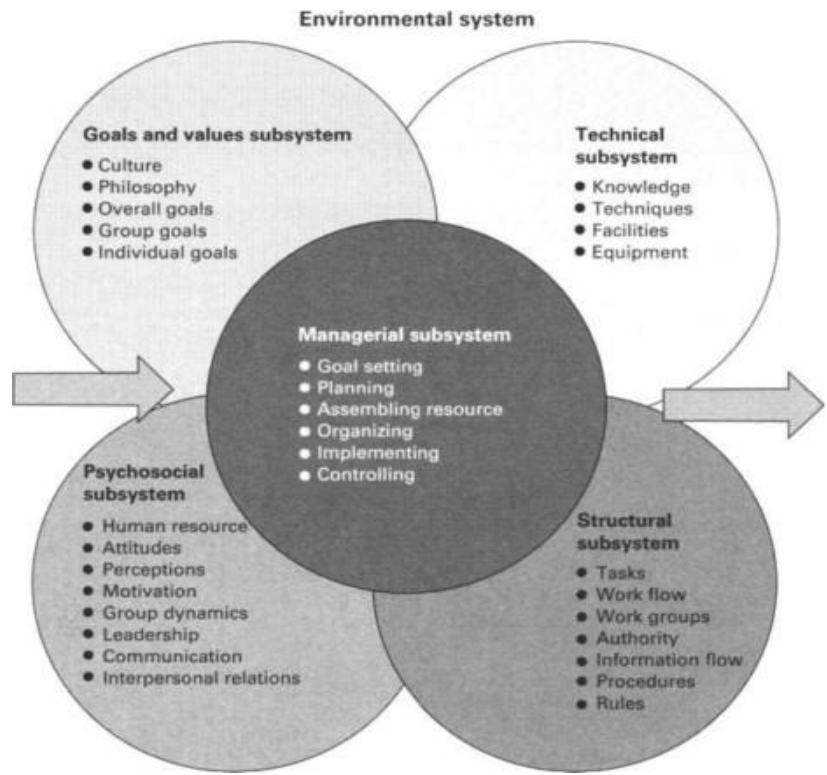

Figure 2: Subsystems of a Socio-technical Systems [15]

A useful idea that offers a common approach is socio-technical systems theory (STS) [16]. This asserts that systems are essentially comprised of three essential and interconnected elements; *technical, social,* and *environmental* (See *Figure 3*). Thus, in addition to the overlap of views between the authors in [15] and [16] about the constituents of a socio-technical system, there is agreement about the existence of interrelationships and interactions amongst system constituents. To re-emphasise the points, it is further noted that some of the key characteristics of open socio-technical systems include; *(i) having interdependent parts, (ii) having an internal environment with separate but interdependent technical and social subsystems, (iii) adapting and trialling goals in external environments, (iv) retaining system goals that can be achieved by more than one means, (v) allowing for design choices during system development, and (vi) the conditions for successful system operations and performance are enabled by the interactions of these factors and more specifically the **"joint optimisation"** of both social and technical factors* [17], [18].

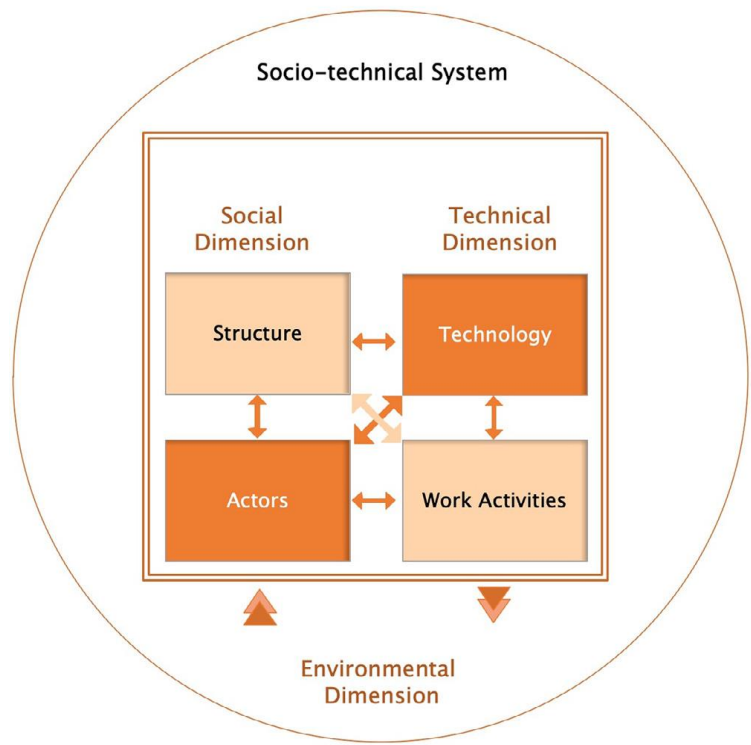

**Figure 3: Socio-technical Systems [16].**

So, with cybersecurity in mind, the traditional definitions of a socio-technical system in [16] in the context of cybersecurity can be stated as comprising;

i. *A security tailored arrangement:* involving people or users with a clear purpose that includes security considerations with either defensive or an adversarial objectives, or both.

ii. *A technical security subsystem working towards attaining and maintaining the security tailored arrangement*: staff and users using security-tailored knowledge, skills, techniques, tools, equipment, and facilities, to realise and uphold defined security objectives.

iii. *A structural security subsystem designed to accomplish the security tailored arrangement:* staff or users working together on integrated activities and processes to realise the defined security objectives.

iv. *A psychosocial security subsystem designed towards accomplishing the security tailored arrangement :* staff or users in social relationships, guided by a security management subsystem to accomplish defined security objectives.

v. *A managerial security subsystem working to accomplish the security tailored arrangement :* a coordinating setup for planning and controlling the overall task for system/organisational security, i.e., ensuring that the activities of the organisation as a whole are coordinated to securely realise organisational objectives.

Putting all these together, a socio-technical security system may be defined as one *'enabling a (cyber) security tailored arrangement using a set of designed and interacting technical, structural, psychosocial, and managerial subsystems and capabilities to help realise defined security objectives'*.

## 1.3 Why 'Socio-Technical' Security?

Understanding the broader compositions of industrial control and cyber-physical infrastructure systems and organisations and combining this knowledge with the lessons from the cybersecurity incidents described earlier, suggests that effective security (for these types of system) is not solely a technical matter. It requires an adequate blend of technical and social viewpoints and measures. Similar to the views about safety within operational organisational environments [19], emerging and thriving operations/process developments means that cybersecurity can be, and is, emerging as a growing feature of organisational life - an enabler for doing and remaining in business. This arises from the integration of control and feedback systems – involving technologies, operations processes, and humans in the loop – and which is degraded by pressures to economise costs and to avoid heavy workloads within an organisational setting [11], [20]. It means that both the technical and social attributes of a system and the associated security risk landscapes all need to be considered at the same time when thinking about security risks, and associated decision-making for design and management [5], [21].

To explain further, the types of smart and interconnected industrial critical infrastructure (cyber-physical) systems that are emerging and being used today are essentially socio-technical. They comprise various technologies co-interacting with themselves and often with humans in the loop to enable the performance of processes leading to achieving certain operational objectives [22]. Essentially, it reflects an interplay of social actors and technical components, demonstrating that it is inappropriate for issues around the security of such a wider system be tied to be technology elements alone. Securing such a system of multi-element interactions cannot be effectively achieved with a measure that does not incorporate factors and behaviours spanning all system elements at the same time.

Focusing on the technical or social perspectives of security alone or in isolation has not been very effective as these often leave gaps in protective capability. Better security can be achieved through deeper considerations of how the technical dimensions of security overlap with the social ones, and how the growing overlap can influence system security or otherwise. This thinking approach emphasises the value of socio-technical security holism in understanding and addressing security threats, vulnerabilities, attack likelihoods, and consequences on modern industrial or cyber-physical infrastructure systems. It also favours the same thinking approach in determining how to effectively control or manage the various risk attributes to ensure the best resilience, hence the least possible impacts and disruptions to societal systems.

## 1.4 The Security Modelling & Simulation Pathway: How the issue/problem can be addressed

Ensuring effective security-thinking for modern industrial or cyber-physical infrastructure systems and operating organisation is vital. Such thinking should broadly consider and address both technical and social (and possibly physical) system elements, attributes, and associated security risks at the same time. However, exploring and experimenting with security approaches directly on live or real-life industrial or cyber-physical infrastructure systems comes with some operational difficulties, including safety and security risks and consequences. Real-system experimentation has been shown to cause potential disruptions to process functions [23]. Nevertheless, if the benefits of modern industrial and cyber-physical infrastructure systems are to be realised optimally, then proper functionality and protection needs to be assured. This means that systems must be safeguarded from disruptions or harm and from any security mal-interventions. Typically, this starts with determining the potential security vulnerabilities and threats, then how to effectively resolve them using appropriate approaches that ensure the best outcome [4].

Modelling and Simulation (M&S) is one common approach adopted by the engineering and computing security community for reasoning about, and exploring security-related objectives in industrial infrastructure systems [24]. Security modelling helps to create a normalised view of security conditions as a representation [25]. Typically, a security model can contain a range of information including details about the network infrastructure, threats, vulnerabilities, potential impacts of attacks, controls, interdependencies, and business services. Thus, a security model provides an effective way to represent the past, current, or future security conditions of a system /organisation. It can be developed to reflect or test defensive capacities – reflecting controls and policies, or adversarial capacities – reflecting areas of vulnerabilities that can be attacked in the architecture under consideration. Security simulation helps to mimic security-related conditions, behaviours, activities, or processes using known information about; a system, an infrastructure, or an organisation's security risk attributes, in a way that corresponds to reality [25].

For modern cyber-physical infrastructure systems, M&S provides focused methods for analysing the behaviours or actions of system constituents including; component dynamics, interdependencies, and cascading effects – all from components interactions [26]. These also need to be modelled and simulated to the extent that is possible to support reasoning about security and resilience, and the impact of interdependencies. M&S uses the results of modelling system constituents, component attributes, functions, and behaviours to support several security analysis viewpoints, and to assist determination or synthesis of appropriate ways to manage security risks.

The construct obtained by considering and integrating both social and technical entities/elements of a system which can influence the system's state of security (and resilience) can be termed a *'socio-technical security model'*. This refers to a security model with the following attributes: *(i) it models and simulates a part or an aspect of a system or organisation's functions, process, or state*, *(ii) it combines two or more of social, technical, physical, or spatial (environmental), and operational attributes of the system or organisation*, and *(iii) it considers either a defensive or an adversarial perspective of security, or both*. These characteristics describe the context of socio-technical security modelling used in this study, and cover approaches that include security analysis software, descriptive methodologies, techniques, procedures, and analytical or conceptual models that inform any aspect of security risk M&S.

This report presents an analysis of available socio-technical security modelling approaches. It forms the first part of the programme of research on understanding the knowledge, perceptions, practices, enabling factors, and barriers to adopting socio-technical security modelling to support threat and vulnerability identification, impact and countermeasure/control analysis and modelling, in industrial and cyber-physical control systems infrastructure (including IoT and associated systems at the periphery of the Internet). More specifically, it focuses on exploring the value and maturity of modelling and simulation in practice, especially for socio-technical security risk analysis of IoT-connected critical industrial infrastructure control/cyber-physical systems.

The report is expected to yield insights on enablers, blockers, principles, and approaches for applying socio-technical security M&S for critical industrial system security. This should provide a usable reference that can support infrastructure system security modellers, researchers, developers, and users in understanding the social, technical, and physical security contexts that can be jointly modelled and simulated, and how this may be achieved. It should lead to recommendations for driving secure-by-design and resilience-by-design in digitalised critical infrastructure systems (e.g., Water and Transportation). The research outcomes are hoped to further support UK government's mission of driving the necessary cultural change in adopting security and resilience design principles and encouraging responsible actions by industry.

## 1.5 Research Questions

This report underscores the development and use of socio-technical security modelling in the critical industrial/cyber-physical infrastructure domains. More precisely, it aims to evaluate the state-of-the-art, application, and maturity of socio-technical security models for industries and sectors dependent on cyber-physical infrastructure systems. It also explores the gap between academic research and industry practices concerning the modelling of the social and technical aspects of security. The research aim will be addressed through the following research questions:

*RQ1.* *What are the common model representation formats/approaches applied to socio-technical security modelling and simulation?*

*RQ2.* *What are the application domains for socio-technical security modelling and simulation?*

*RQ3.* *What are the maturity levels of the existing approaches for socio-technical security modelling and simulation? (How widely used are existing socio-technical security modelling and security approaches?)*

*RQ4.* *What is the geographical spread of interests in socio-technical security modelling and simulation?*

*RQ5.* *What socio-technical security contexts are modelled? (elements/attributes of socio-technical systems that are modelled, security dimensions of focus, aspects of security modelled and simulated, scope of security covered)*

*RQ6.* *Is dependency context/attribute considered?*

## 2. Methodology

This section describes how the research was conducted. It covers descriptions including; the definition of research, how the relevant literature was identified and gathered from articles, references, databases and the internet, how the articles were examined to identify relevant literature related to socio-technical security modelling and/or simulation, and the criteria used to guide the selection of the most relevant sample of articles, the extraction of relevant data based on defined review criteria, and the analysis and reporting of findings. *Figure 4* presents a visual description of the process followed.

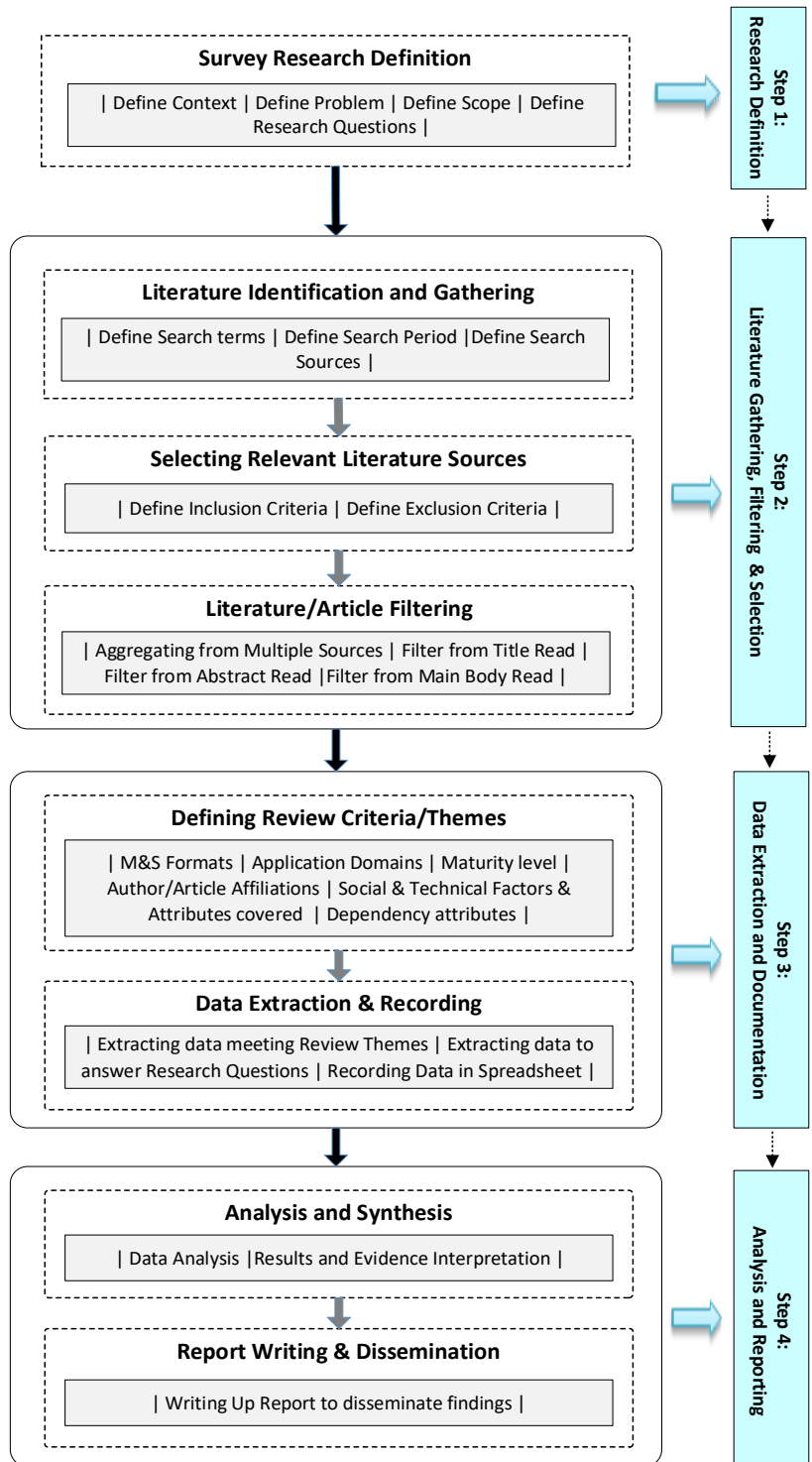

**Figure 4: Study Research Methodology**

## 2.1 Literature Gathering

To identify the relevant literature, appropriate boundaries of context were defined to direct the literature gathering process, and to ensure that an appropriate scope of work and results specifications were maintained. The following search constrictions were used:

i. Articles and Literature reported in the English Language, which is the official language within the environment of study.

ii. Articles or reports on theoretical developments and/or applications of socio-technical security modelling and simulation – security modelling and simulation analysis, methods, tools, and techniques that considered and combined technical and social elements of a system. This helps to answer the research context defined in the project aim.

iii. Theoretical or practical socio-technical security modelling and simulation works that are related to critical infrastructure domain use cases (applicability to systems in energy, transport, water and wastewater, finance, communications, health, dams, government system, etc.,). This helps to address the research scope defined in the project aim.

iv. Research articles or reports that have academic, industry, or government affiliates or support.

## 2.2 Search Sources and Search Terms

The study commenced with initial searches and aggregation of literature from relevant article sources. For this, SCOPUS, and Web of Science (WoS) article databases were used. These were selected because each of them covers a broad and multi-disciplinary range of article reference sources, including popular science, engineering, and social science domains and communities such as; IEEE, IET, ACM, Springer-Link, Elsevier (ScienceDirect), PLOS ONE, Taylor & Francis, Sage, etc. Essentially, SCOPUS or WoS is considered a database of databases. Thus, both databases were selected because jointly they enable access to more resources covering a multidisciplinary (socio-technical) domain; implying the strength of a wider coverage and resource concurrency [27]. Search phrases use Boolean combinations of keywords: 'Socio-technical Security', 'Modelling', and 'Simulation' or 'Analysis'. The search period was from 2000 to 2020 for the relevant literature used to draw primary data and analysis for this study.

Horizon-scanning of how developments around socio-technical security thinking have progressed within the literature domain was also explored. This involved using the same keywords as the primary data search within both SCOPUS and WoS databases to identify the occurrence of related articles dating from 1900 – 2021. Such high-level data may be useful to improve the understanding of how socio-technical security concepts have evolved and progressed over time, and what might be learnt.

It is typical to find an article appearing in multiple databases during independent searches. Often, this adds the extra task of further investigation to identify the originating source (article database) of the publication. This is effort and time-intensive in the overall literature search, filtering, and gathering. The use of SCOPUS and WoS article reference databases brings the benefit of identifying and choosing a single article source regardless of multiple occurrences, thereby reducing the likelihood of a large number of redundant article results. This saves time and effort during literature gathering process.

In addition, similar searches were performed on the Google search engine using similar keywords and phrases listed above. This was done to broaden the outcomes of the study outside of the academic domain and to avoid missing relevant works on socio-technical security modelling that may be useful to the security modelling community.

## 2.3   Article Filtering and Context Data Extraction

To arrive at the initial number of relevant papers to be analysed based on the search terms used, the number of articles obtained from SCOPUS and WoS were combined. A number of articles were found in both the SCOPUS and WoS lists. To maintain a single instance of each article, duplicates were removed. This resulted in a refined list of relevant articles with only single instances.

Further filtering out of articles was carried out based on the inclusion and exclusion criteria earlier described. This was done by inspecting each article's title in the first step (Level 1 filtering), reading through each article's abstract in the second step (level 2 filtering), and where necessary skimming through text in the article's body in the third step (level 3 filtering) to identify and select the articles that met the required criteria and to exclude those that did not. This resulted in a further refined list most relevant sample of articles for the primary study. *Figure 5* Shows the article filtering and selection process.

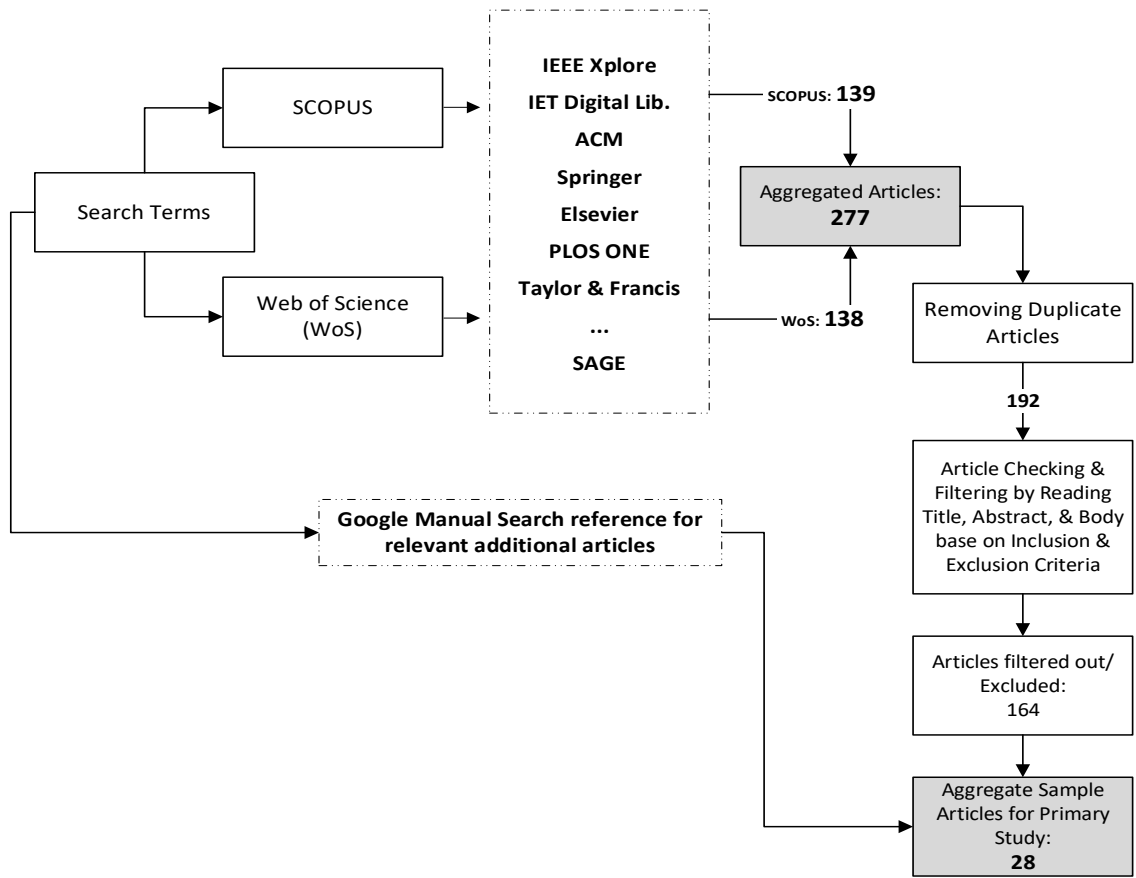

**Figure 5: Research Literature Filtering Process and Final Outcome**

To further the study, relevant contextual data and information were extracted from the final sample of literature articles that satisfied the outlined inclusion and exclusion criteria. Data extracted from each article was based on the review themes and attributes described in ***Sub-section 2.4.*** Specifically, this relates to information on the socio-technical security M&S which helps to answer the research questions through informing attributes concerning; model representation formats, application domains, maturity level (use context), work/author affiliations, socio-technical attributes covered, security contexts and attributes covered, and dependency characteristic coverage. The data extracted was recorded in an Excel spreadsheet to facilitate analysis.

## 2.4 Review Criteria for **Evaluating** Socio-technical Security M&S Approaches and Rationale

Based on knowledge of the various constructs that socio-technical security models can take, together with the outlined research questions, the following contexts are used to analyse the literature selected on socio-technical security M&S:

1. Modelling & Simulation formats and approaches for socio-technical security.
2. Application domains for socio-technical security models.
3. Maturity levels of socio-technical security models in terms of real-world use.
4. Affiliation of Authors/Works on Socio-Technical Security Modelling.
5. Security contexts and attributes covered by socio-technical security models.
6. Dependency attributes considered in socio-technical security models.

Review criteria such as 'critical infrastructure application domains' and 'dependency attributes' have been used in prior studies around understanding the responsiveness of existing critical infrastructure protection approaches in the light of the adoption of emerging technologies such as IoT [4], [28], [29]. These have led to useful insights and recommendations, and provide inspiration and a basis for using similar criteria in this study.

### 2.4.1 Modelling & Simulation Formats and Approaches for Socio-Technical Security

This refers to the structural components or building blocks used to represent socio-technical security contexts in existing works. Similar to the technical dimensions of security modelling [30], socio-technical security M&S can be described by these categories; methodologies [31], techniques, conceptual frameworks, languages (schemes) [32], and artefacts (tools in hardware or software forms), and the analysis and/or implementation of procedural guidelines. Each of these can manifest a different level of maturity in the development and use of the socio-technical security idea or context. Often, categorisation aims to simplify or speed up the characterisation and implementation of infrastructure systems or use cases. Socio-technical security can apply to various critical infrastructure domains, and often includes various dimensions and contexts of security.

A socio-technical ***security conceptual framework*** will consist of a set of pre-designed procedures, processes, and language structures that enable a foundation for reasoning about, and building a socio-technical security application or use case. Following the same context, a ***language*** describes a specified method for defining or communicating the social and technical security interactions or risk attributes within a critical infrastructure use case, and in a form that is understandable to particular readerships. Typically, a **_methodology_** describes a way of doing something. It includes a method associated with an underpinning idea, concept, and philosophy [31]. Relative to the study context, it should present parameters for addressing a socio-technical security risk reasoning problem via M&S. Similar to frameworks, a methodology can be expanded to include specific components such as phases, tasks, methods, techniques and tools [31]. A **'tool'** refers to any instrument (digital, conceptual, or physical, or all three) that can be used to perform certain functions or operations. Relative to the study context, a tool will typically encompass a defined procedure and language for representing a joint social and technical security interaction or risk scenario. A socio-technical **'application'** represents the practical use or acting out a socio-technical security modelling method, technique, procedure, framework, or a tool that has been previously learned. '**Analyses**' further extends the practical use by assessing and identifying the key elements that comprise or affect the socio-technical security application or results, using the appropriate skills to address the central problem.

To help draw inference about the common M&S formats and approaches for Socio-Technical Security, for each of the relevant articles selected, we extracted and analysed specific data about the associated model's; *'classification' , 'representation format', 'sub-representation format (where available)', and 'modelling technique'*. Identifying the various M&S constructs that apply to socio-technical security can support understanding of the extent of suitability and applicability of certain constructs to particular contexts of socio-technical security. In addition, this also provides an opportunity to identify and highlight the limitations of each security modelling and simulation format, together with an understanding of how improvements and extensions toward better maturity can be explored. This is helpful when combining multiple formats or constructs to achieve better and more useful security representations.

*2.4.2 Application Domains of Socio-technical Security*

This refers to the various real-world system domains where the concept of socio-technical security modelling and simulation are tried or applied. This is considered in the context of critical infrastructure sectors. The sectors in this category are termed 'critical' because of the importance of the functionality they provide or support in maintaining normal societal operations. Another reason is because of the harmful consequences that can happen from their failure due to any form of disruption or destruction. In this study, a harmonised list of critical national infrastructure (CNI) sectors is drawn from definitions in the Revised National Infrastructure Protection Plan (NIPP) of the US [33], the European Union Directive 114/08 [34], and the UK CPNI Documentation [35]. This results in a total of fifteen CNI sectors considered in the study. These include Energy (electricity, oil, natural gas), Transportation (Railways, Roads, Highways, Aviation, Shipping and Ports), Water and Wastewater, Chemical, Industrial Control, Dams, Defence Industries, Emergency Services, Financial Services, Food and Agriculture, Government facilities, Commercial Services, Health and Public Health, Information Technology and Telecommunication, and Nuclear.

Where necessary, the analysis also extends to the micro-unbundling and characterisation of CNI sectors and their associated use cases. Thus, to help draw inference about the common application domains where socio-technical security modelling is being applied, for each of the relevant articles selected, we extracted and analysed specific data about the *'critical infrastructure sector'*, *'the sub-infrastructure domain (where indicated)'*, and *'the functional/operational service use case'* where the associated security M&S approach is applied. This helped to characterise and identify, to a reasonably low-level, the specific real-life application contexts and aspects where socio-technical security M&S is being considered or applied. Furthermore, it helped to highlight the specific system/sector functionality, while exploring the concept of socio-technical security, how it is being applied, and the lessons to be drawn, especially for sectors and domains not currently adopting the approach.

*2.4.3 Maturity levels of Socio-Technical Security Model*

Maturity is used to imply the level of robustness, standardisation, widespread acknowledgement and/or adoption of a socio-technical security modelling idea or construct within both academy and industry. A similar criterion has been used in prior studies [36] to evaluate tools and techniques for critical infrastructure protection. To draw insights about the maturity of the socio-technical security approach, details about the *'application extent'* and associated *'citation count'* of each concept were extracted and analysed.

Regarding the extent of application, it is possible to gauge the maturity of socio-technical security models by considering the state and the progressive development of a certain M&S concepts found or acknowledged in available literature, from procedural outlines through to artefacts. Concepts and models can be theoretical, proofs-of-concept, or extended to practicable use within certain application environments. These could typically be a test environment if the concepts are still going through evaluation and assurance processes, or a real-world systems environment if qualification and accreditation are achieved.

Another indicator of maturity involves acknowledgement which can be drawn from the number of times a specific modelling ideas or concept (with related literature) is cited by others. Technically referred to as citation counts, this attribute offers a way to gauge or construe the relative scientific significance or perceived quality of papers [37], [38], or impact of an author, an idea, or a publication. Relative to the study context, this can help to understand the level of viability for adopting an identified or highlighted socio-technical security modelling approach based on evidence provided by scientific critiques, reviews, and modifications/improvements to the underlying theory. This can further help with the easy categorisation of concepts/works according to adoption viability, and identification of those with the most promising outcomes and prospects.

*2.4.4 Affiliation of Authors/Works on Socio-Technical Security Modelling.*

Author affiliations refer to the formal organisational associations that relate to the socio-technical security M&S work under review, or that relates to the authors of the literature. This information is often used to rank institutions based on the number of articles their faculties and researchers have published in certain areas [39], [40]. It can also point to the geographical distribution of the sources of research work in an area. To infer common author affiliations for the works on socio-technical security M&S, we extracted data about the *'Author Affiliations'* for all the relevant literature selected. That is, the organisation or institution the author belongs to, or identifies with. This helps with highlighting the relevant research experts, groups, and clusters that focus on socio-technical security modelling and simulations, and their detailed areas of interest.

This information can be combined with information from other criteria such as *'domain application'* and *'citation count'* to strengthen a view about the viability and maturity of certain socio-technical security M&S concepts. The relevance and use of M&S concepts within the scientific community can be further highlighted, thus pointing to potential places – institutions and clusters – where research/work collaborations might be explored.

*2.4.5 Security Contexts and Attributes in Socio-technical Approach*

Security contexts and attributes refer to the various aspects, characteristics and dimensions considered in reviewed socio-technical security M&S literature. These include the social and technical elements and attributes considered in the M&S concepts. This heading also includes the scope of M&S and security dimensions covered.

To make inferences about common socio-technical contexts, we inspected each work in the sample information that is indicative of the broader *'socio-technical factors included'*, and *'specific social and technical system attributes covered'*. Also, to infer about common security contexts, we extracted from each work in the sample information that was indicative of the *'security dimension of focus'*, *'security contexts modelled'*, and *'security scope covered'*. By security dimension, we mean the overarching goal or approach of a security initiative.

From a high-level objective of security modelling (and simulation), security contexts and dimensions can be explored from either *protective* or *adversarial* pathways [41]. A protective view is also referred to as *'defensive security'*, and the adversarial or attack view is referred to as *'offensive security'* [42]. Defensive security describes security measures (technology, policies, and/or procedures) intended to withstand or prevent attacks or avoid cyber-related risks (vulnerabilities, threats, attacks/sabotages). Offensive security describes measures intended to break or sabotage a system to either test the strength or robustness of systems and identify potential vulnerabilities [43], or damage the system purposely. Penetration test harnesses and 'red-teaming' are examples of offensive security, used to test the defences and resilience of a target system.

Typically, a security discourse, action, or counteraction tends towards one or the other of the defensive or adversarial directions or combines both. This categorisation offers insights into the leading objectives of interest to security modellers, analysts, and system owners, as well as possible motivation for socio-technical security research activity. A breakdown of contexts into associated attributes of socio-technical factors and the functional security aspects, can also offer insights into areas of growing security concerns, possible knowledge gaps and associated interests of the academic and industrial communities. It also helps to underscore how socio-technical security thinking and developments are being explored, while opening an opportunity to identify gaps in existing approaches guided by global security standards and best practices for security risk management.

## 2.4.6 Dependency Attribute Covered

This involves exploring the interactions between socio-technical system elements and their effects on security. Multi-element compositions of industrial/cyber-physical critical infrastructure systems including social and technical attributes and their interconnectedness can mean that functions and services are facilitated and improved due to interactions and interdependencies. However, it also means that actions that exploit susceptibilities and result in negative consequences can enable such consequences to cascade through interconnected critical infrastructure subsystems. Thus, we checked each literature in the study sample to find if *'dependency attribute(s) is(are) considered or not'* in the modelling.

It may be helpful to consider if and how system interdependencies can, or are being explored from a socio-technical security perspective to gain the benefits of achieving a more secure system. Several damaging cascades in critical infrastructure security incidences [44]–[46] are consequences of interdependent relationships amongst the broader system constituents. This in turn can contribute to the degrading of system security, safety and dependability [47].

## 3. Results and Analysis

In this section the results of the analysis of selected literature that meet the review criteria discussed earlier are presented.

### 3.1 Year Distribution of Literature by Year

#### 3.1.1 Horizon *Scanning* Analysis

Initial horizon scanning of the progressive developments around socio-technical security thinking, analysis, modelling, and/or simulations drawing from the selected literature reference databases SCOPUS and WoS presents quite interesting results as shown in **Figure 6.** The oldest instance of a related article was found in SCOPUS in 1984. Between that time and 2021, 139 articles were found in SCOPUS and 138 articles in WoS. Articles relating to socio-technical security started emerging more predictably in 2003 for WoS and 2006 for SCOPUS. Before these times, there were few scattered occurrences of related articles in both databases. A gradual increase in the number of articles was observed from 2003 for WoS and from 2007 for SCOPUS. More articles appeared in both databases in the last decade (2011-2020 with 122 articles for WoS, and 124 articles for SCOPUS) than in the decade before (2001-2010 with 14 articles for WoS, and 13 articles for SCOPUS) which probably suggest that socio-technical security thinking only started gaining weight and interest in the last decade. Though the yearly distribution of articles varied between the two databases, the total number of articles in each of the database is not too dissimilar. This suggests some level of consistency in pattern and trends on how interests and efforts towards socio-technical security have evolved and might be progressing. Although the results may not reflect the totality of all available literature in publication databases globally in the domain of study, the result offers an idea about the potential totality of articles.

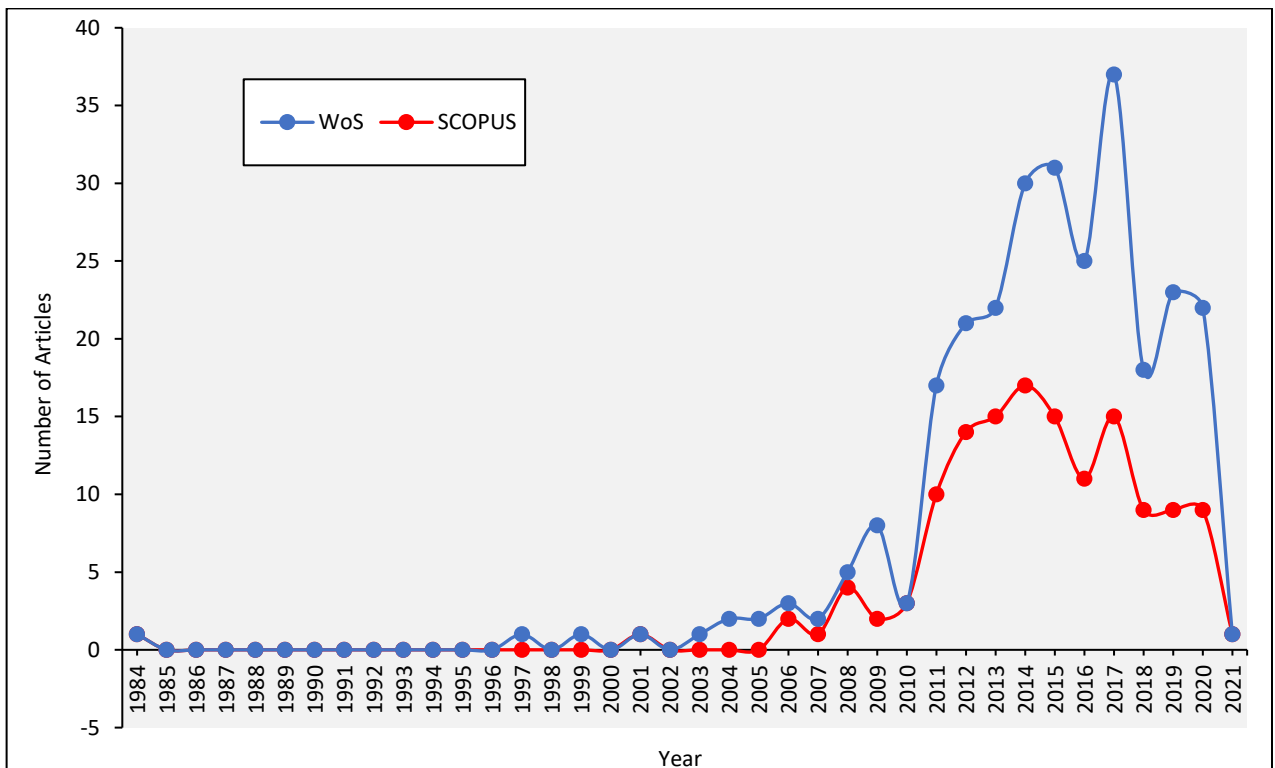

**Figure 6: Related Article Occurrences in SCOPUS and WoS on Socio-Technical Security**

*3.1.2 Primary Search Analysis*

Following the literature gathering, merging, and article filtering processes based on the defined inclusion and exclusion criteria, a total of 28 articles were found to be relevant to the context of this study. See Appendix A for the outline of sample literature. This represented the sample literature for primary study and spanned the period between 2004 and 2020 (See Error! Reference source not found.). The primary literature sample obtained may not also capture the totality of resources available covering the study context, however, given the outcome of the horizon scanning, we believe that this represents a good sample from which relevant insights and conclusions may be drawn. This number of articles suggests that the idea about reasoning and modelling security considering both technical and social attributes equally, is not entirely recent as it would seem – literature suggests a couple of interests and conceptualisation going back a decade ago. However, emphasis appears to be growing stronger in recent times. More literature on socio-technical security modelling and its significance for improved security reasoning appears to be emerging, as the emphasis on multi- and cross-disciplinary research and development continues to rise with many references to collaborations between sciences and the humanities.

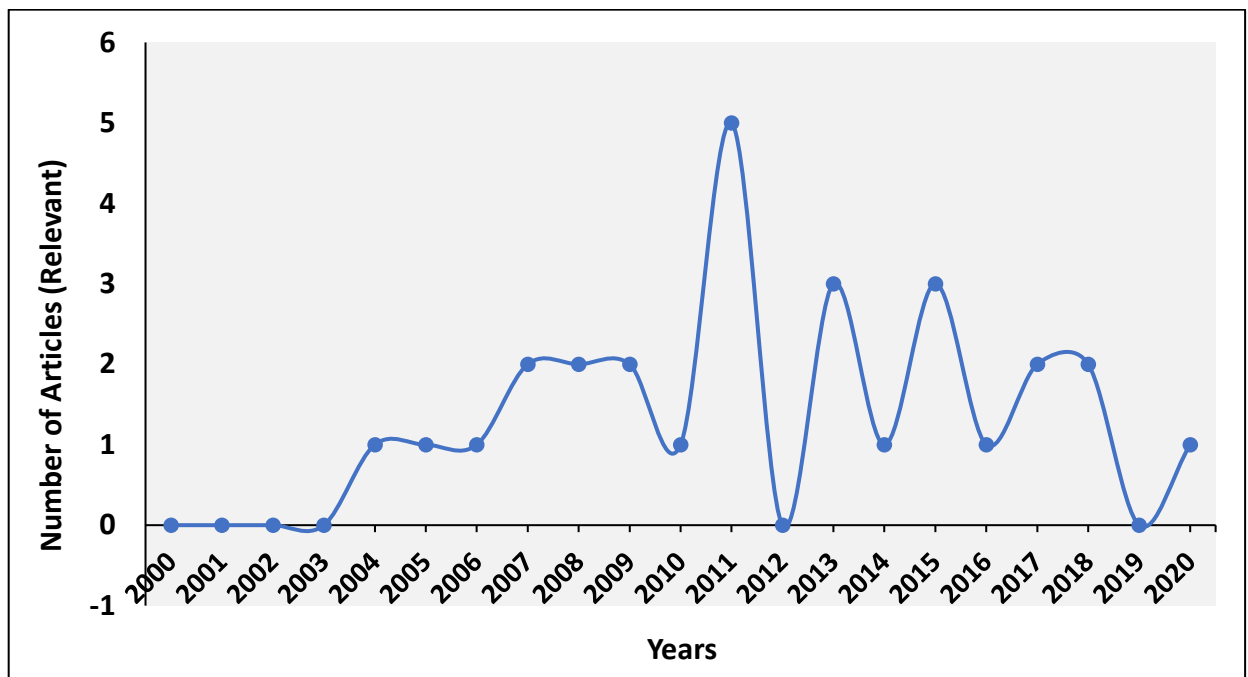

**Figure 7: Distribution of Primary Literature Sample for Study on Socio-Technical Security**

## 3.2 Modelling Techniques & Representation Format

The works reviewed adopted a variety of model representation techniques and formats ranging from frameworks and methodologies to formal language definitions, software tools, analyses, and application contexts. From *Figure 8*, results show that some of the works appear to combine one or more model representation/development formats. At least one work combined/captured two or more model development forms. For example, Zhou et al [48] proposed a 3D security modelling platform that can capture and model security requirements in a Social IoT environment. This modelling included a methodology and a tool that is a graphical notation-based extension to the Business Process Modelling Notation (BPMN). The work by Probst et al [49] on *'Insider Attack*

*Modelling'* included a methodology, a language and analyses, while Aslanyan et al's [50] work on modelling and analysis of a socio-technical system encompassed a methodology, an analysis, and a description of an application context.

From ***Figure 9***, the *'Methodology'* model representation format; either used alone or in combination with other model formats appeared most times (17 instances). This represented the most common model representation format that authors used to characterise and/or convey their thoughts about their socio-technical security modelling concepts. Another common model descriptor that followed is the use of 'Analyses' alone or in combination with other formats to characterise a socio-technical security context. This appeared 10 times from the sample. Other common representation formats include Tool and Language formats.

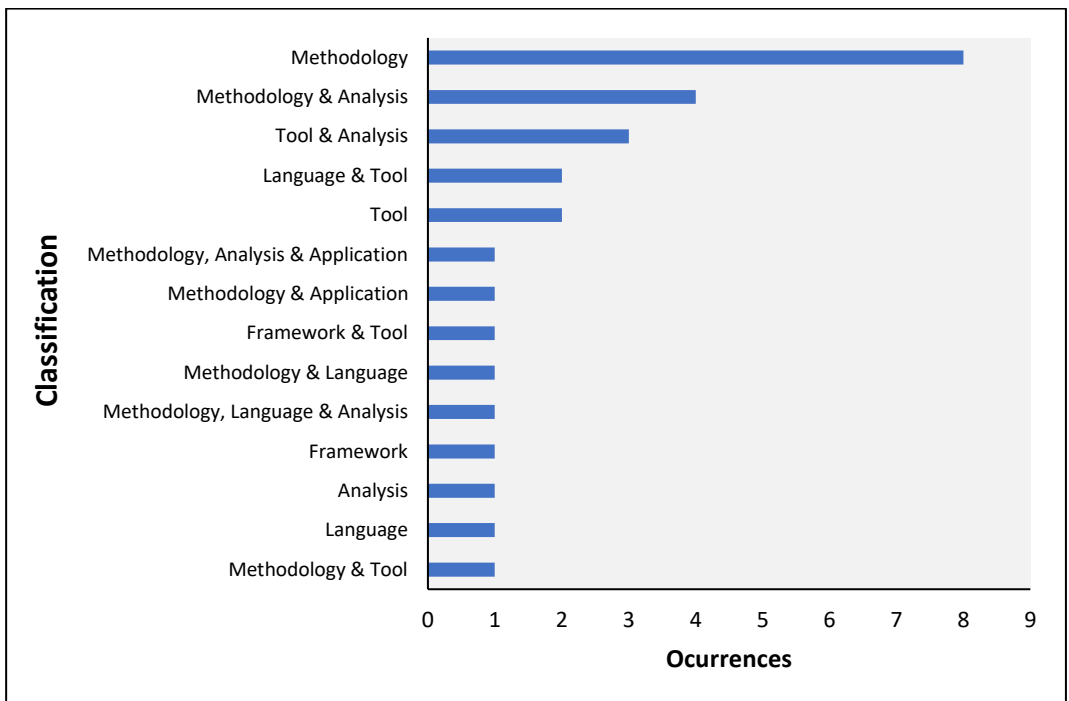

**Figure 8: Occurrences of Modelling Techniques and Representation Format**

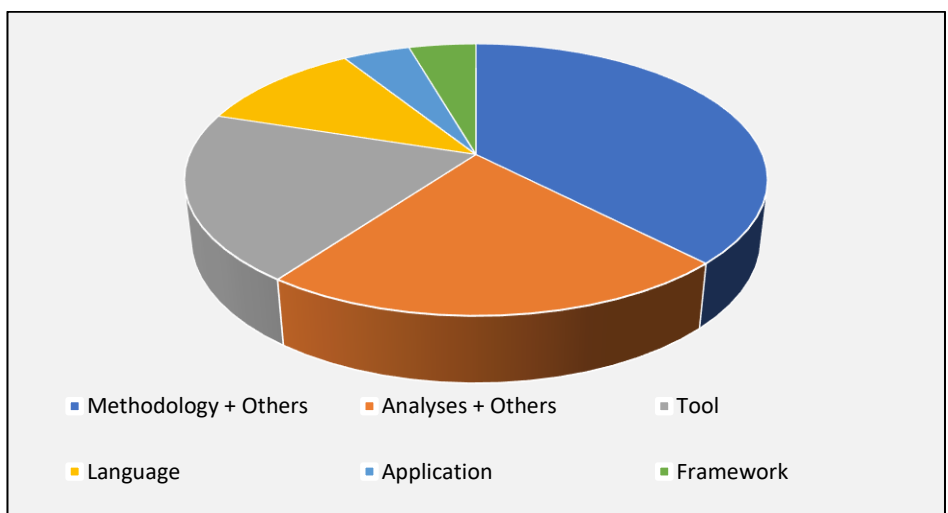

**Figure 9: Common Model Representation Formats**

In terms of structural representation, from **Table 1**, diagrammatic structures appear to be more commonly used for modelling or illustrating socio-technical security contexts. They appeared 13 times in the sample, and took on a range of specialised forms including; Unified Modelling Language (UML) in [51], Data Flow Diagrams (DFDs) in [52], Business Process Management Notation (BPMN) [48], and simple Plane Shapes combined with texts and/or with mathematical formulations [25], [53]–[56]. Formal graph notation was also used by some authors [57]–[59], and in one instance was combined with Tree Notations [60], algorithms [61] and/or maths equations [49].

Table 1: Model Structural Representation & Occurrences

| Sub Representation formats | Occurrences |
|---|---|
| Graph | 6 |
| Textual | 1 |
| Graph + (Trees or Algorithms or Maths Equation or Tool) | 7 |
| Maths + Diagrams +Textual | 1 |
| Plane Diagrams | 1 |
| Plane Diagrams + Text Descriptions | 8 |
| UML | 1 |
| DFD | 1 |
| Trees | 1 |
| BPMN | 1 |

### 3.3 Application Domain for Socio-Technical Security Modelling Approaches

In the UK, there are 13 named critical national infrastructure (CNI) sectors. These include; Transport, Water, Chemicals, Energy, Civil Nuclear, Communications, Defence, Emergency Services, Finance, Food, Government, Health, and Space [35]. However, the list of CNI sectors covered in this study includes a compilation of CNI sectors from the UK CPNI documentation [35], the European Union Directive 114/08 [34], and the Revised US National Infrastructure Protection Plan: NIPP [33].

We note that some infrastructure sectors have been classified as critical in the US, but do not appear in the UK CNI list. These include *Commercial facilities*, *Dams*, and *Critical Manufacturing*. The UK Communications sector appears to include both the Information and Telecommunications sectors; these were separated in the US CNI list. The growing interconnectedness and interdependency of CNIs to ensure a functional and smooth-running society is clear, as is the negative effect of cascading failure impacts that cause disruption to civil society. Interdependent CNIs suggests a need to rethink what constitutes a CNI in the UK, but this is outside the scope of this study. From the literature, a harmonised list of CNI sectors was obtained and used to evaluate the application domain for socio-technical security modelling approaches. These include: *Transportation, (Rail, Roads, Waterways, Air, Shipping and Ports), Energy (electricity, oil, gas), Chemical, Manufacturing, Dams, Defense Industries, Emergency Services, Financial Services, Food and Agriculture, Government facilities,*

*Space, Commercial Services, Health and Public Health, Water and Wastewater, Communications (Information Technology and Telecommunication), Nuclear.*

Table 2: Application Domains and Distributions for Socio-Technical Security Modelling Approaches

| CNI Sectors | Occurrences | Sub-Domains | Occurrences | Application Areas/Use cases |
|---|---|---|---|---|
| Chemical | - | - | - | |
| Commercial Facilities | - | - | - | |
| Manufacturing | - | - | - | |
| Dams | - | - | - | |
| Defence Industries | - | - | - | |
| Emergency Services | - | - | - | |
| Energy | 1 | Enterprise Systems architecture modelling for Electric Power Utility | 1 | Attack Probability estimation in enterprise system of Electric Power Plant |
| Financial Services | 3 | Secure Online Banking System | 1 | Usable and secure online payment system |
| | | Secure Automated Payment Processing System | 2 | Banking Payment Processing system |
| Food & Agriculture | - | - | - | |
| Space | - | - | - | |
| Government Facilities | 4 | E-Government Business Process & Information Exchange Security | 4 | Security requirements for trust and secure exchange in E-government software systems for Publica Administration |
| Health & Public Health | 1 | Identifying, modelling, and analysing the security of health information. | 1 | Security requirement analysis for secure software development. |
| Transportation Systems | 1 | Transport Tourism | 1 | |
| Water & Wastewater | 0 | - | - | |
| Nuclear | 0 | - | - | |
| Communications (IT & Telecommunications) | 5 | Information Technology | 2 | Security of Enterprise systems |
| | | Telecommunications | 3 | Ubiquitous and Mobile Computing |
| Generic | 13 | Cyber-physical/Industrial Control Systems | 5 | Cyber-Physical System Security Use Case |
| | | Generic Digital System | 4 | Digital Forensics & IoT Use Cases |
| | | Enterprise/Organisation-wide System | 3 | Automated Security Risk Assessment |
| | | Physical Intrusion Security Model in Building Systems | 1 | security requirement definition for physical building structures: An Intruder malignity model |

From **Table 2**, out of the 16 compiled CNI sectors used, 5 sectors classified by UK CPNI as 'critical' had at least one reference application or a use case for socio-technical security modelling. These include Communications (IT & Telecommunications), Government facilities, Financial Services, Transportation, Health/Public Health, and Energy (Electricity) infrastructure sectors.

IT & Telecommunications had the highest singular sector occurrence of use cases where socio-technical security M&S was applied with 17.86% of the works reviewed (i.e., 5 out of 28 use cases). Specific sub-areas/scopes for M&S applications included ubiquitous and mobile computing, and general enterprise systems. Other singular sectors with instances of applications include Government facility services with 14.29% of reviewed works, and Banking/Financial Services with 10.71% of reviewed works. The Government facility applications sub-areas were in E-government business process management & Information Exchange Security. These involved exploring specific use cases around trust and secure digital transactions amongst multiple parties in e-government platforms for public administration. The banking/financial services sub-areas of applications were in banking transaction processing with specific use cases around secure and usable payment processing systems considering both technical and social attributes.

Nearly half (46.43%) of the reviewed works on socio-technical security modelling seemingly applied to more generic domains which employ industrial control and cyber-physical systems, and Information Technology (IT). Some application use-cases for socio-technical security modelling included IoT and explored security-specific contexts around automated security risk assessment, the usability of security, security requirement definition, and digital forensics.

### 3.4 Maturity of Socio-Technical Security Modelling Approaches

To determine the maturity of current socio-technical security modelling approaches under review, we used citation data related to each work. By citation, we imply the number of times other research has referred to a specific M&S work. In addition, we sought for statements within each work that suggests instances or areas where the socio-technical approach is applied either in theory or in practice, as well as the type of system, sector or environment involved.

Having a number of citations may suggest a growing recognition and/or acknowledgment of a specific security modelling and/or simulation approach. It can also highlight the potential of a progressive scrutiny and improvement approach in certain security M&S, with multiple instances of both theoretical analysis and practical scenario testing leading to applications in real-world systems. An approach that demonstrates such progressive scrutiny and evolution can potentially yield better maturity than one with less or without. It is important to understand the maturity of security M&S tools to help the community assess the capabilities, strengths, and limitations of various approaches. This can minimise the chances of adopting half-baked security approaches and solutions. With maturity, a security M&S approach or tool might be said to have endured rigorous progressive development involving critiques, reviews, modifications, optimisations, and hardening.

Maturity reduces the risks of failures and enhances performance, productivity, reliability, and trustworthy outcomes.

*3.4.1 Citation Results*

Regarding the citation numbers for the reviewed works, the socio-technical security M&S work by [53] appears to be the most cited with 219 citations at the time of reporting this study. The work was first published in 2007 by authors from the University of East London (UK) and the University of Trento (Italy). The authors presented the *'Secure Tropos'* methodology *(an extension of the traditional Tropos methodology)* for reasoning and modelling security requirements in a software system development process. The work presented a *proof-of-concept theory* illustrated using a health and social care application domain.

A further investigation into the approach showed that the *'Tropos'* concept proposed by the same authors dates back to 2002 [62]. However, the concept development appears to have progressed over time through the exploration of test applications [63] thence evolving into a *language* and a modelling *tool* referred to as the *'ST-Tool'* [54], [64], [65]. Knowing about this evolution offers some insight concerning the level of maturity of the idea and framework underpinning *'Secure Tropos'*. Although the initial work referenced focused on demonstrating security requirement modelling for 'health and social care software system development process', we also found other reported applications of the *'Secure Tropos'* method, language, and tool to telecommunication, travel planning and organisation, air traffic management, and e-government operations use cases [66]–[69]. We also found other related works that referred to the *'Secure Tropos'* methodology, the STS-ml tool, or related applications to certain use case environments. Summing up the citations from various related works linking to the central *'Secure Tropos'* socio-technical security M&S concept results in about 267 citations. This further highlights the growing recognition and acknowledgment of the *'Secure Tropos'* methodology and approach for socio-technical security requirements engineering in the research community.

Another socio-technical security modelling approach that has significant recognition through its citation count is by the authors in [51], presenting a language and tool called *'CySeMoL'* for modelling enterprise-level systems architecture and assessing vulnerabilities. At the time of reporting, the work had 59 citations. It includes a reference to a practical application of the tool to model the enterprise system segment of a real power grid's control system. The authors were from the Royal Institute of Technology, Stockholm (Sweden).

There were other works with more citations, for example; the *'Portunes'* security modelling framework [70] for modelling attack scenarios considering physical, digital and social attributes had 36 citations. *'Portunes'* used the *KLAIM* [71] family of languages to model interactions and apply to systems with physical, digital, and social components generically. Additionally, there was the work by Pieters [58] which also had 36 citations, this was another methodology and analysis approach for modelling the actions of humans for information security. It used the *Actor-Network Theory* to model human actions, and was applied to cyber-physical building security access management systems. Both works represented *proofs-of-concepts*, and their authors came from the University of Twente in the Netherlands.

*3.4.2 Application Extent Results*

Regarding the extent of applications of the socio-technical security modelling approaches reviewed, 27 (representing 96%) of the approaches appeared to present *theoretical proofs-of-concept* as shown in **Figure 10**. These involved demonstrating security modelling concepts or ideas to show that they work. Essentially, each work appears to highlight how it incorporates both social and technical attributes to support the reasoning about a specific security context. In cyber or information security, *proofs-of-concept* often suggest initial phases of system development intended to minimise functionality risks while improving the confidence that the prescribed approach adequately addresses the use case. Only one (i.e., 4%) of the reviewed socio-technical approaches stated a real-world application in practice; a real national electricity power utility station in Sweden [51]. This is the *'CySeMoL'* security modelling language and tool. The expectation is that this specific tool has undergone reasonable testing and scrutiny to an assurance level sufficient to allow its practical application in a highly critical national infrastructure setup. Although there is no clear evidence of this, we observe that at the time of reporting this study, the *'CySeMoL'* language and tool for socio-technical security modelling was one of the highly cited works amongst those reviewed.

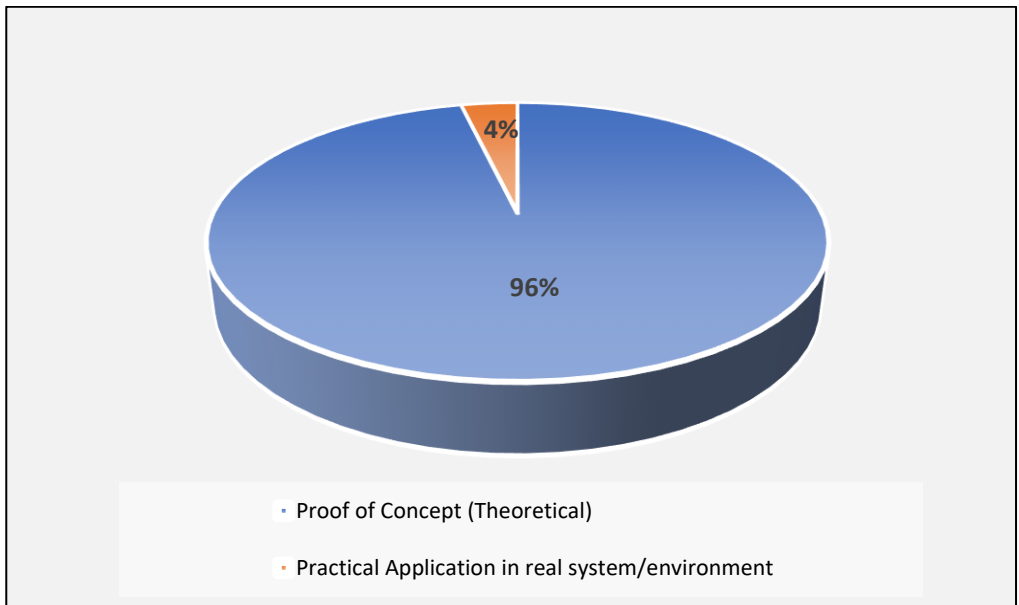

**Figure 10: Extent of Application**

## 3.5     Affiliation of Authors of Socio-Technical Security Modelling Works

Concerning author affiliations, geographically and institutionally, 6 of the literature exhibits (approximately 21.43%) came from authors at the Technical University of Denmark as shown in **Table 3**. The University of Twente and University of Trento had 4 (i.e., 14.23% of the sample) items each, and University of Cambridge has 3 items (i.e., 10.71%). Other institutional affiliates that had fewer literature exhibits are shown in **Table 3**.

Concerning the categorisation of sources based on countries, *Figure 11* shows a wider literature spread in the United Kingdom and Denmark with each having a total of 6 (representing 21.43% of the sample) of the works reviewed. These are followed by Italy with 5 items (17.86% of the sample) and the Netherlands with 4 (14.29%) of the sample literature. These results present a revealing insight about the geography (countries) and domain demographics (learning domain and disciplines) where the thinking about socio-technical security modelling and simulations is gaining weight, and potentially being researched.

Table 3: Institution-based Affiliation of Authors of Socio-Technical Modelling and Simulation Approaches

| Affiliates | Countries | Occur | Percent |
| --- | --- | --- | --- |
| University of Luxembourg (1) | Luxembourg | 1 | 3.57 |
| Norwegian University of Science and Technology (1) | Norway | 1 | 3.57 |
| University of South Africa (1) | South Africa | 1 | 3.57 |
| The State University of New York at Buffalos (1), Salient Works (Industry - 1) | USA | 2 | 7.14 |
| Stockholm University (1), Royal Institute of Technology, Stockholm (1) | Sweden | 2 | 7.14 |
| University of Twente (4) | Netherlands | 4 | 14.29 |
| University of Trento (4), Politecnico di Milano (1) | Italy | 5 | 17.86 |
| University of Cambridge (3), University of East London (2), Liverpool John Moores University (1), Middlesex University (1), University of Brighton (1) | United Kingdom | 6 | 21.43 |
| Technical University of Denmark (6), Aalborg University (1) | Denmark | 6 | 21.43 |

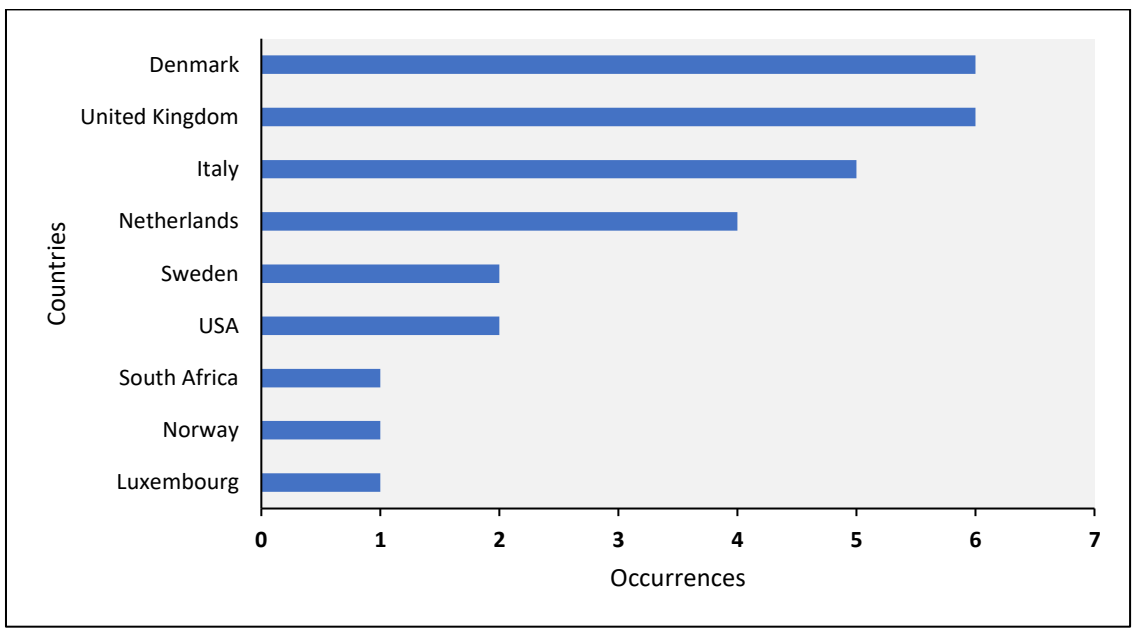

Figure 11: Country-based Affiliation of Authors of Socio-Technical Modelling and Simulation Approaches

## 3.6 Security Context Covered by Socio-Technical Security Modelling works

There are different security areas and contexts that appeared to have been considered and modelled in the literature reviewed, and this also covered a range of social and technical factors that affect security as presented below.

### 3.6.1 Socio-technical system elements and attributes modelled

For the social and technical factors and attributes considered in various security M&S works reviewed, nearly 82% included attributes that relate to human agents or actors that reflect social factors/dimensions (See *Figure 12* ).These include actor/agent goals and their ability or tendency to reach system components, actor/agent roles and tasks, skills, motivations, behaviours, attitudes, perceptions, and knowledge, all in relation to security. Nearly 79% of reviewed works considered security policy attributes linked to levels of exposure definitions for human actors/agents, as well as digital and physical entities. These policies included authorisation rules, restrictions on access to digital and physical assets and locations, and actions of actors/agents. Nearly 18% of reviewed works considered attributes linked to organisational structures. Specifically, these referred to processes related to information-sharing, adoption of security strategies, security policies and standards. Nearly 7% considered culture and value attributes related to organisational behaviours. Nearly 4% considered relationship and interaction attributes such as connections of locations, resource sharing, ownership of assets, and containment of network(s) and device(s) locations.

For technical factors/dimensions, *Figure 12* shows that about 68% of works considered technology attributes encompassing both digital and physical forms. Nearly 46% of works considered resources and assets from tangible and intangible perspectives.

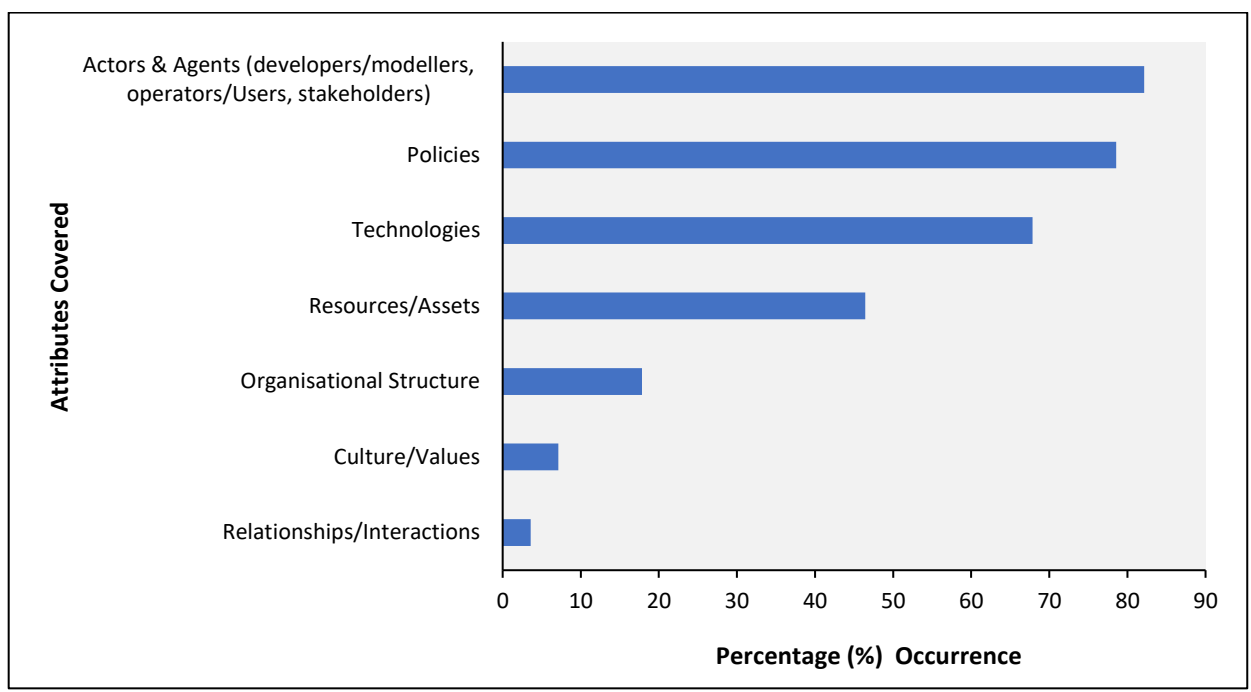

**Figure 12: Modelled Socio-Technical System elements**

### 3.6.2 Security Dimensions

For security dimensions as shown in *Figure 13*, results show that 15 (54%) of the reviewed works focused on *adversarial security alone*. By comparison, defensive security focuses on reactive security measures that are concerned with prevention, detection, and response. This is often viewed as the conventional type of security which includes deploying security tools such as anti-malware and anti-intrusion to help identify and deter malicious events before they happen. We found that 11 (39%) of the reviewed works focused on *defensive security alone*. This type of dimension for security tends to focus on the proactive measures of seeking out potential malicious actors/agents or attributes in a system, and in some cases attempting to weaken or disable their capacities to harm. For example, the measure of identifying areas of weaknesses in components and networked systems, and taking steps to close up such vulnerabilities. Often, these can start with some adversarial measures such as penetration testing a network or web application. We found that 2 (7%) of the reviewed works involved a *combination of adversarial and defensive security*. As expected, this combines both defensive and offensive approaches to achieve a broader scope of security. In this sense, the capacity to prevent, detect, and respond to security risks is complimented by an ability to anticipate potential risks and prepare for them.

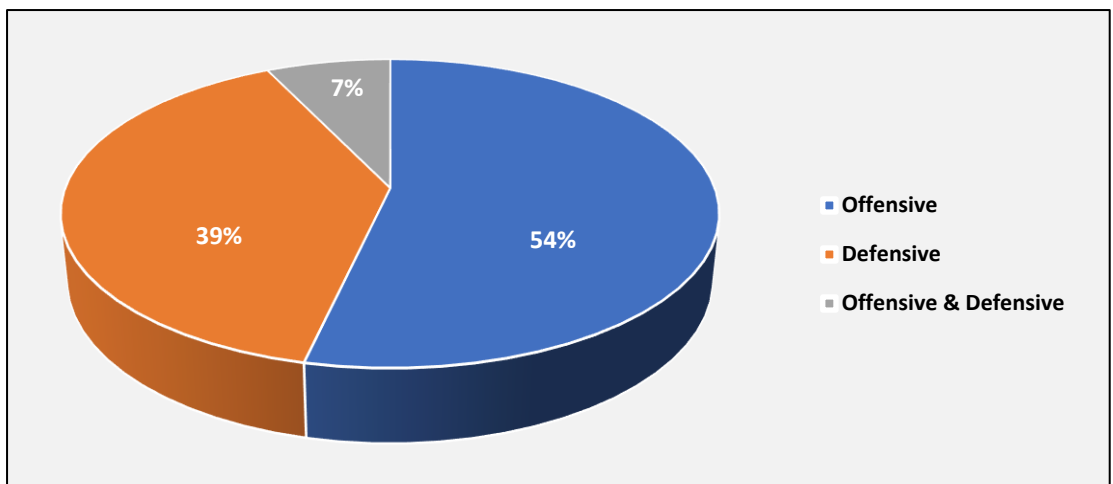

**Figure 13: Security Dimension**

### 3.6.3 Security context/aspects modelled

Regarding the security contexts and aspects modelled in the socio-technical security works, 12% focused on modelling security attributes around the control of unauthorised exposure of data and/or information (See *Figure 14*). Also 9% concentrated on reasoning about security requirements for software-based systems, 18% focused on resolving security control and countermeasure attributes for technology infrastructures, 3% looked at modelling human or personnel-level security needs, 6% focused on modelling organisational-level security needs, and 9% concentrated on modelling security requirements around system component interactions. Furthermore, 31% of reviewed works also considered or included attributes related to attack(er) strategies, and access to system resources in their modelling, while 12% also considered the reachability of system components and the characteristics of attacks and their perpetrators. A grouping of the security contexts modelled in the sample is also presented in *Table 4*.

Table 4: Security Context and Characteristics Modelled

| Security Attributes that are Being Modelled | Occurrences | Literature (STS-Model List) |
|---|---|---|
| **Threats to the Security & Privacy of Communications Data/Information** | **4** | |
| • (The Control of unauthorised exposure of Data/Information) | 4 | 1, 2, 3, 23 |
| **Security design requirements for secure system design, development, implementation, and operations.** | **15** | |
| • (Security requirements for software system) | 3 | 4, 13, 25 |
| • (Security Control/Countermeasure features for tech Infrastructure) | 6 | 5, 6, 11, 15, 26, 27 |
| • (Human/Personnel security needs) | 1 | 12 |
| • (Organisational security needs) | 2 | 16, 18 |
| • (System component interaction needs) | 3 | 17, 24, 28 |
| **Component interactions & Attack Strategies/attributes-based resource access and/or reachability to system components** | **14** | |
| • (Attack(er) strategies and resource access) | 10 | 7, 8, 9, 10, 11, 13, 14, 23, 24, 26 |
| • (Attack(er) Attributes/characterisation and reachability) | 4 | 19, 20, 21, 22 |

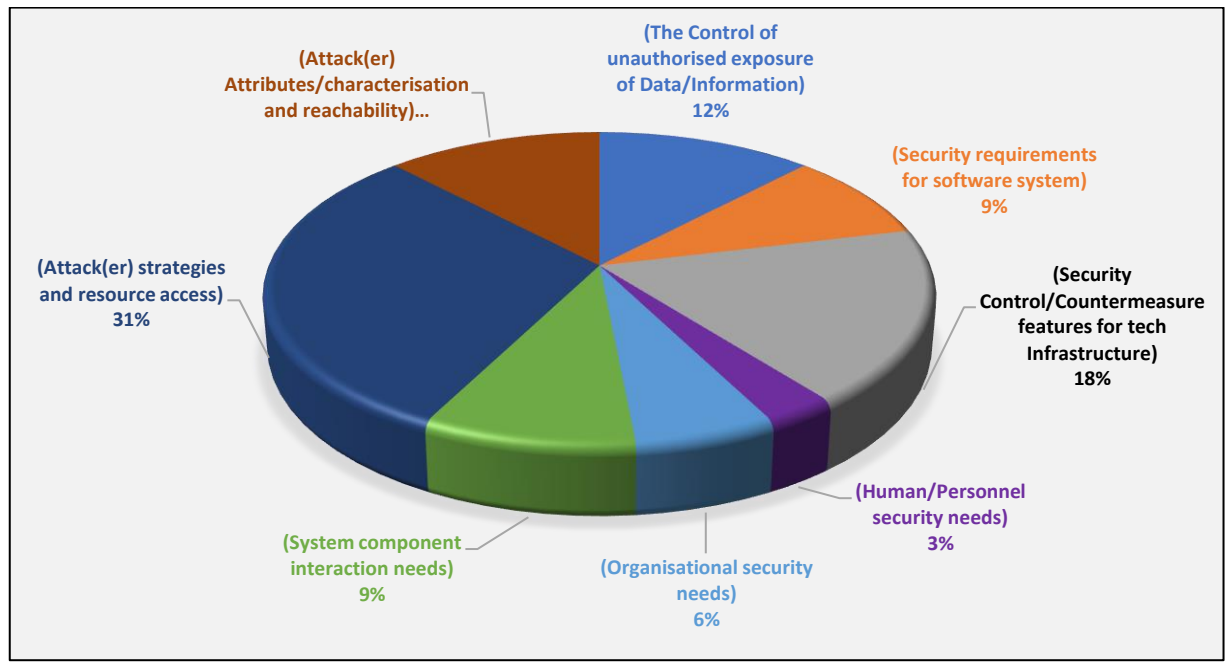

Figure 14: Security Characteristics modelled?

## 6. Dependency Attribute Considered (Results)

Concerning whether dependency attributes are considered in the sample works on socio-technical security modelling, results shown in *Figure 15* indicate that only 8 (representing 29%) considered or included some attribute of dependency or interaction in their set-up, while 20 (representing 71%) did not include any attribute related to dependency.

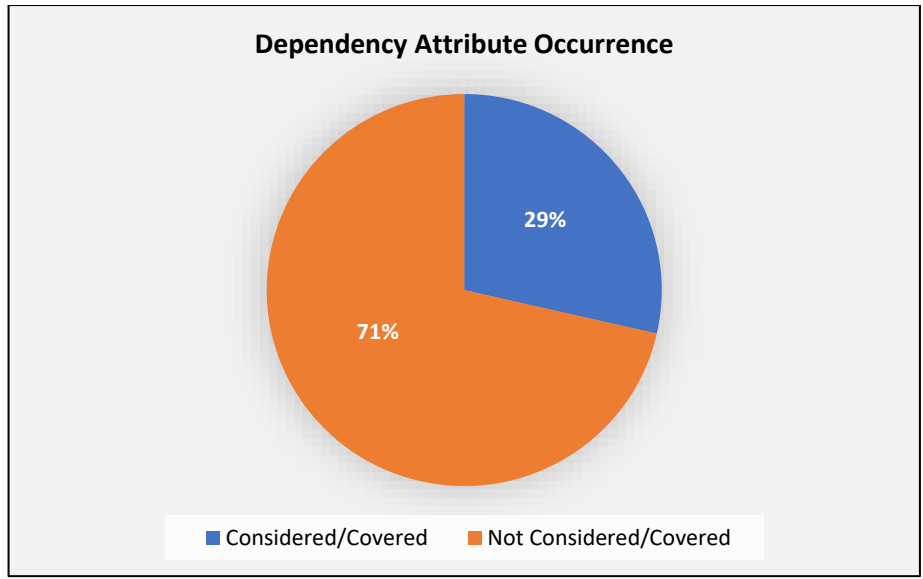

**Figure 15: Sample distribution for Dependency Attribute Covered**

## 4. Discussion.

In analysing the results presented, we draw useful insights related to the objectives and contexts studied.

### 4.1 Distribution of Literature (by year)

On the emergence and distribution of literature, both the horizon scanning, and primary search indicate a gradual rise in the number of items of literature exploring socio-technical security thinking and M&S over the period considered. This suggests a steady growth in the emphasis placed on socio-technical security approaches. It is useful to explore the reasons for this. Firstly, it could be due to growing support for adopting socio-technical system design (STSD) approaches which advocate design methods and processes that consider both social (non-technical) and technical factors, which influence the functionality and usage of digital systems in the workplace. This might have been further prompted by the gradually growing recognition that digital systems and workplaces do not only comprise technologies, but also include non-technical (social), human, and sometimes physical elements. Moreover, all these elements contribute functionality and co-interact for the system to function properly. Secondly, there is the gradual change in the nature of cyber threat and attack landscape of digital infrastructure systems. Attack vector directions appear to be changing from technical to non-technical targets and modes [6]. Existing developed and mature technical security solutions do not necessarily provide sufficient capacity to effectively ensure protection from new and emerging variants of cyber-attacks and compromises.

Thus, this may be driving the growing acknowledgement that new ways are needed to address such growing problems. These new approaches may include exploring and adopting a systematic and constructive use of combined social and technical principles and methods in analysis, design, testing, evaluation, operation, and evolution of complex systems. If CNIs involve complex interactions between humans, technologies, and the environmental aspects of operations, then a solely technical security solution approach would appear limited. A socio-technical approach to reasoning about operations and security can provide a better way to address the growing dynamics in issues rather than the narrower approach of using technical or social viewpoints separately. Clearly, there is a growing and wider recognition and interest in the socio-technical approach within the last decade than in the decade before the last. This is expected to continue to grow as new trends emerge and better understanding is gained on the implications of the interactions amongst social, technical, and environmental factors of systems; helping to ensure that the security risks and impacts of such interactions become clearer.

### 4.2 Modelling Techniques and Representation Format

Regarding model representation formats, it is worth highlighting the frequency of 'methodology' approaches in the reviewed works. Often, methodologies tend to hinge on, or relate to, preceding ideas or concepts which in this study is referenced by; *'how security modelling and simulation have been commonly conceived or approached in the past'*. References frequently point to the more common approaches for creating or representing models of the interrelationships and interactions amongst technical security components and their attributes in digital systems. This is good in the socio-technical security context, as it highlights the understanding and effort to not just articulate the concepts on how socio-technical security can be modelled and/or simulated, but to do so by building on underlying theories. This is essentially using supporting philosophy to describe how the method - including both social and technical system security-related factors and attributes - can be modelled to reason about security risks, their impacts, or how they can be effectively managed.

Only a smaller proportion of the methodologies reviewed appear to widen their scope to include language definitions, developed tools, or descriptions and analyses of theoretical or practical applications. This suggests that current works in this area are still in early development. Also, it means that most current research initiatives and efforts on socio-technical security M&S – especially related to cyber-physical or IoT-based industrial systems – are still exploring and/or establishing the theoretical base for understanding various ways effective security can be achieved and developing best practice use cases. More clarity and confidence in results and outcomes are needed to progress these works to develop tools, real applications, and analysis. The practicalities of how these may be achieved still need to be established with tests and demonstrations that can provide proof that theoretical concepts, modelling propositions and simulations actually work, and can mirror reality.

The common use of diagrammatic representations using graphs, data flow diagrams, business process management networks, tree structures, and unified modelling language structures, further highlights the view that socio-technical security M&S is in its early development stages. Furthermore, these representations highlight efforts towards simplifying the understanding of systems by using visualisation notations to show how the elements involved in socio-technical systems behave, connect, and co-interact. Diagrammatic notations help in promoting a deeper understanding of socio-technical systems, and have value in initial security requirements definitions and generation. This approach differs from the solely technical, which appears to have progressed towards more robust, complex, and often automated design tools (software and hardware), which are continuously undergoing reviews and upgrades.

Therefore, it would be desirable and beneficial to see existing socio-technical security modelling diagrams progress further towards the level of maturity seen in technical security modelling. This would mean being able to achieve system security M&S, tools, techniques, and approaches that can capture complex socio-technical behaviours, interactions, and interdependencies while reasoning practically about security contexts. This could take the form of tools (software/hardware) or other integrated productivity enhancements that speed up the implementation of process objectives, thereby influencing better, timelier, and more proactive security understanding, decision-making, and management response(s). Notwithstanding, it might be difficult to have a single socio-technical approach – framework, tools, techniques, analyses, or methodologies – that can provide a *'fit-for-all'* solution for addressing all security risks, resilience, dependencies and impacts in the cyber-physical control systems used in CNIs. This has also proved difficult in purely technical security M&S contexts and approaches [72]. However, it would be useful to understand how any two or more approaches may be combined to achieve a wider span of socio-technical security applications.

Typically, this involves drawing from the pool of known or existing security M&S constructs in the knowledge domain. Sometimes, it would involve new formats that build on existing ones, either by adding new attributes to existing constructs or combining multiple existing constructs to arrive at a potentially improved approach or schema. But the complexity involved in certain socio-technical security relationships e.g., ones that enable a vulnerability between a technology attribute and the attribute of an organisational structure or policy, may make certain security modelling formats unusable or difficult to represent in applications. Thus, it would be more rewarding in terms of effort, resources, the effectiveness of results, and reliability if the commonly applicable and easily usable security M&S formats and constructs were used. Currently, the common approaches for socio-technical security M&S appear to be in the form of methodologies; describing how security may be perceived and the associated parameters represented. Some methodologies appear in expanded forms, including different combinations of tasks, phases, methods, language, and tools, often leveraging the common modelling forms such as system dynamics, agent-based, or empirical-based with their respective benefits.

### 4.3 Application Domain for Socio-Technical Security Modelling Approaches

Concerning the application domain for socio-technical security M&S approaches reviewed, results indicate the critical infrastructure sectors that are predominantly exploring the socio-technical security modelling approach to address a range of objectives. This highlights pointers to critical operational domains and professional communities where increasing concerns, discussions, and efforts to evolve solutions are emerging regarding socio-technical cybersecurity. These often relate to the risks, frequencies, dependencies, and potential direct and cascading impacts. Information Technology and Telecommunications, Financial Services, Government facilities, Health and Public Health, and Transportation appear to be the sectors that are more active and gaining momentum in considering and exploring a socio-technical dimension of security. It is not likely that cybersecurity measures do not already exist within these sectors, but more likely that existing security measures are not sufficiently effective or holistic to prevent or mitigate cybersecurity incidents.

These findings also highlight the inactivity and potential laxity in other critical sectors, and the need for greater efforts towards understanding the socio-technical dynamics, risks, and implications of effectively securely operating those critical sectors. Also, they highlight the need to adopt socio-technical security thinking and M&S approaches to address inherent and emerging security issues in order to safeguard the sectors and wider society. This is especially desirable for sectors categorised as 'critical' by the UK government, such as energy, water and wastewater, where interests and efforts towards socio-technical security thinking (including M&S) appear to be minimal or absent.

The leading focus on the IT and Telecommunications sector is apparent. Works that claimed more generic sector applications covered applicable security use cases in Industrial Control Systems (ICS), Internet of Things (IoT), digital forensics, mobile and ubiquitous computing, and organisation-wide automated risk assessments. However, this still suggests close links to IT and Telecommunications. Specific socio-technical application contexts or references covered appeared to refer to, or link to the IT/Telecommunication parts of these systems. This emerging interest in a socio-technical approach to security may be associated with the rising cases of cyber compromises/incidents targeting IT/Telecommunications, other affected sectors, and the application areas described. Other potential reasons accounting for the rising view and effort in socio-technical aspects of security may include; the persistent increase in the numbers of successful cyber incidents affecting businesses and organisations, the growing realisation of how cyber-attack directions are evolving, i.e., breaching digital/electronic systems within organisations by targeting and exploiting non-technical (users and other social) system elements, the gradual acknowledgement that technical security solutions alone are inadequate to protect critical infrastructure system assets, and the desire for security M&S approaches that can address both the social and technical vulnerability attributes of critical infrastructure systems at the same time and equally.

In government or the public sector where there is increasing global interest in digitalising governance, e.g., via e-government capabilities, to improve governance efficiency and productivity, a socio-technical security approach is clearly crucial to secure such government services. In the UK, e-government platforms and services must comply with government security standards and protocols in relation to appropriate security controls for people, processes, and technologies – network infrastructure, data, services, and the cloud. This is necessary since a key category of stakeholders in e-government platforms typically includes users of electronic government systems and platforms who are unskilled and uninformed in security, and who are at the risk of intentionally or unintentionally interacting with technology and processes in a way that can enable security vulnerabilities, or successful cyber compromises. A similar scenario applies to the Banking/Financial services sector where customer electronic payment transactions and processing services are involved. Most successful cybersecurity breaches affecting this sector appear to have some socio-technical dimensions and have often led to huge financial and reputational losses to individuals, organisations, and financial institutions.

Without doubt, the integration of technology innovations such as internet, IoT, Big Data, AI, etc., into critical infrastructure sectors enables performance, productivity, and economic benefits. However, these often add to the structural and operation complexity of critical infrastructure technology footprints and usability, which further widens the CNI cybersecurity risk landscape. The security risks are both technical and social and require a similar combination of measures and countermeasures to enable a far-reaching solution. Critical infrastructure operators that are embracing IT innovations must be aware of these, and purposefully adopt a socio-technical security thinking mindset while exploring their solutions.

## 4.4 Maturity for Socio-Technical Security Modelling Approaches

In using citation counts and modes/levels of application to infer the maturity of existing socio-technical security approaches, it might be thought that a security model should be presumed 'mature' when it has many, or wider, acknowledgements and/or endorsements through some form of verification, validation, and/or accreditation. Also, this might be assumed if such a security model has got to a state in its development lifecycle that it is used or usable, in real systems/environments to bring clear benefits to addressing real security issues. Models with fewer citation counts can suggest little awareness, evaluation, and traction which may be indicative of lower maturity, especially if available in public domain.

While citation counts offer a useful insight towards work or model maturity, it is still limited in a way. A body of work or security model can be cited multiple times by original or different authors, yet such multiple citations might not necessarily emphasise the same valued context or the same level of understanding. So, it might be difficult to establish a strong case for the maturity of a model using wide citations without clarifying that all the citations emphasise or highlight the same utility of the cited approach for achieving the same or similar objective. Rigorous context-specific validation of the utility and dependability of a security model is more useful than just drawing from secondary or tertiary thoughts from the referenced works to advance a different idea. A high number of citations serves better to show a wider recognition of a cited work, security model, or a related idea (s), and perhaps suggests that more people are considering or exploring the idea. It does not necessarily mean that the idea is being improved, thus it does not provide the strong support needed to indicate effectiveness and dependability. This should be considered when drawing conclusions about model maturity.

Notwithstanding, the 'STS-ml' and its associated methodology (Secure Tropos) appear to be much cited. Also, it seems that this language and methodology can be applied to several system formations for information or control modelling. This is so far as the formation comprises interactions amongst humans (actors or agents) organisations, software and/or hardware, and involves the transmission/exchange of data or information. The *'Secure-Tropos-based'* STS-ml tool's development appears to have progressed to address the difficulty in the preceding version in effectively mapping high-level security concepts to low-level security requirements to support improving technical security mechanisms. The newest version of the tool achieves this in a logical and easily understandable way by enabling an interface to drive the process of software requirement engineering (SRE) for a domain of interest. This would start from defining high-level system abstractions through to a process of refinement to obtain specific and appropriate security requirements that can be integrated into security mechanisms.

Regarding application extent, results show that the majority of existing works around socio-technical security modelling for critical infrastructure systems are still in their theoretical conception and analysis stages at the time of this study. These do not appear sufficiently matured to be applied to real/live system environments. This aligns with the insight from prior work [18] suggesting that socio-technical security modelling approaches are in their initial developments and similar to other security-related schemes, are typically following the security product development life cycle. The rate at which social factors are being successfully exploited within industrial domains is increasing and alarming. The devastating consequences and impacts of socially engineered security compromises are to be avoided. By now, it might be expected that mature socio-technical security M&S methods or tools would be available that can realistically support security risk management process development and implementation. With the current state of socio-technical security models, it would seem that the stage for real-life system applications is still a distance away, and researchers in this space need to keep pace with the dynamics of the threat landscape.

Nonetheless, the situation is not entirely irredeemable, as it reflects a mindset and understanding of the need to get socio-technical security modelling right at the idea development and testing stages, and to allow ample scrutiny and validation before security tools are moved to the real CNI environment. The proof-of-concept stage enables the extensive analysis and clear definition of security requirements. It offers a platform and opportunity to build strong and convincing value propositions for socio-technical security M&S schemes, leading to wider acknowledgement and traction.

### 4.5 Affiliation of works on Socio-Technical Security Modelling

For author geographical origins and affiliates, results show the countries and institutional clusters where works have happened or are ongoing on socio-technical security thinking, modelling and simulation. Work in this area appears to be gaining more traction in Europe with stronger interests and involvements in the United Kingdom, Denmark, Italy, and the Netherlands. These countries showed a cluster(s) of researchers exploring cyber/information security thinking that considers and combines both technical and social factors and attributes simultaneously or equally.

Note that most of the works reviewed seem to be created within academic institutions, and this may involve collaborations with industry, the extent of which is not clear. Some of the works seem to be supported through research programmes funded by government, and only one work clearly showed the involvement of a private industry – Salient Works in the USA [73]. An example of government support is for works from the University of Twente supported by the Research Sentinels Programme (www.sentinels.nl) and financed by Technology Foundation STW, the Netherlands Organisation for Scientific Research (NWO), and the Dutch Ministry of Economic Affairs [74]. Another example is from the University of Trento which is funded by the European Union Seventh Framework Programme [22], [68]. Thoughtfully, having most critical infrastructure systems owned and operated by private organisations and combining this with the broader harmful impacts of CNI attacks should strongly motivate greater private sector involvement in security research and development investments for the purpose of improving security, resilience, and safety. Private sectors owners and operators of CNIs would often be in the front line of organisations impacted by any successful cyber-attacks/intrusions, and the first to feel whatever resulting harm that may emerge.

A closer look at the affiliates of authors of reviewed literature reveals a diversity of backgrounds and disciplines. Institutional academic departments spanned Computing and Technologies Institutes, Computer Science and Engineering, and Computer Laboratories. Research group names span various computing and technology areas including Informatics and Modelling, Information and Communications Technology, and Industrial Control Systems. Author disciplines appear to cut across Computer Science, Informatics and Mathematics, System Safety and Security Analysis, Design, Modelling and Simulation (M&S).

The dominance of the computer science and engineering disciplines amongst authors exploring socio-technical modelling and simulation fits well with the institutional affiliates of authors as identified. Also, this may be linked to the view that prior to the trend and focus on cybersecurity at the intersection between computing and engineering, typically, both disciplines separately make widespread use of M&S techniques, skills, and processes for research investigations and solution developments within and across their domain. Thus, M&S is not entirely new to both disciplines.

Contrary to the expectation raised by the term 'socio-technical security modelling' – implying the involvement of both potentially technical and social science experts – social science domain experts appear to be under-represented in the author composition of the works reviewed. This might be due to a potential lack of common usage of the terms *'modelling'* and/or *'simulation'* by social scientists. In their typical work environments and/or activities, social science researchers rarely engage with processes that relate to these technical terms. Another reason could relate to the lack of understanding or interest by social scientists in considering or exploring the influence and/or implications of technical security contexts on social factors around security risks. This aligns with our initial views on how security is typically being viewed and addressed, i.e., the focus being largely on technical aspects for issues and solutions, and isolated from the social challenges and solutions. This can mean potential limitations in the thought-synthesis, analysis, and development of potential solutions, to effectively address emerging security challenges linked to the usage of new technologies such as IoT, Edge Computing, and Artificial Intelligence, etc., in socio-technical industrial domains.

## 4.6   Security Context Covered by Socio-Technical Security Modelling works

For socio-technical system elements and attributes modelled, results show that both technical and social factors and attributes are being considered in the design and development of security for infrastructure systems. The factors found can be categorised under several headings including *technologies, actors or agents, relationships and interactions, policies, cultures and values, resources or assets available, and organisational structures*. Each factor can be decomposed further to outline some related attributes represented in different ways and forms within security models to improve the development of security. For example, technologies are crucial factors which can include (i) controlled plant/equipment (physical), (ii) digital hardware (computers, IoT, etc.), and (iii) Code running on the digital hardware (software/digital) attributes. These attributes work together to establish or define desirable security states. Also, results show that *actors* or *agents* can refer to *humans*, characterising *developers, operators, users*, or *attackers. Agents* or *actors* can also refer to *program builds that are capable of acting autonomously*. *Actor/agent* sub-attributes that can make or mar security can include *roles/responsibilities of actors/agents, their target goals or objectives, their motivations for acting, resources accessible to actors/agents, competencies – knowledge and skills for humans*; and *knowledgebase for programs, perceptions, behaviours,* and *attitudes of actors/agents and related costs*. If any of these properties are not considered, or if considered in the wrong frame or proportion, they can enable security weaknesses capable of causing a system to be less secure and more exploitable. In the appropriate frame or proportion, this can also mean that a system can become less vulnerable and more difficult for malicious actors and actions to succeed.

*Security policy* attributes refer to rules for activities, processes, and states that help to ensure the desired type/level of security within a system or organisation. Often, security policies apply to the *system* or *organisational assets* – human, physical, and/or digital – and can exist to address organisational security, system security, or a specific security context. For example, security policies around defining levels of security exposure, transparency, and/or authorisations for all types of assets within the system were common occurrences in reviewed works. Also found were policies around access restrictions on assets and locations for both physical and digital instances. Also frequent were policy instances around appropriate competencies/capabilities and actions for human actors/agents.

Generally, *security policies* (*and procedures*) are used to set rules for actor/user behaviours, establish baseline security states to minimise cybersecurity risks, define and enforce consequences for security violations, and enforce proper compliance to regulations. They are there to help address potential security risk attributes, i.e., threats, vulnerabilities, and impacts, guide the implementation of strategies for mitigating the security risk attributes, as well as guide on how to recover when exposed in part or full to the security risks. *Technologies* – hardware and/or software – are the common mediums for bringing security policies and procedures to life within socio-technical systems and organisations. These offer a capability to have direct contact or interaction with security risk attributes, enabling them to be minimised and revealing how they affect normal or stable system functions. Technologies also enable the set-up and monitoring of organisational security best practices, and assurance of adequate security compliance among others. So, to be effective or to yield the desired result, security policies need to be clearly defined, understood, and followed. It might be easier to enforce security compliance than to incorporate security policies into security models and mechanisms to capture the desired notion of security within a system or infrastructure. The latter means that security policies are translated and integrated into a broader security reasoning, schema, and/or technologies. The extent to which this can be achieved remains an open question. However, what would be interesting would be to understand how to intentionally incorporate various policy attributes into an overarching security reasoning, security design, and security modelling such that they enable a capability to deter the occurrence of; an exploitable vulnerability, a threat activity or incident, and/or reduce any associated/potential consequences and impacts on real environment functions.

*Organisational structure* attributes in the reviewed literature which appeared relevant in security include *processes around information sharing*, and *the implementation of security strategies, policies, and standards*. *Culture* and *value* attributes still link to the behaviour of actors. Common *relationships* and *interactions* attributes highlighted include *connections*, *resource exchange/sharing, position of actors, possession of assets/resources*, and *containment of network and device locations* amongst others. *Resources/Assets* can be categorised as *tangible* which still refer to physical assets or hardware when thinking from a technology perspective, or *intangible*, i.e., non-physical and/or software -type assets.

There are a couple of upfront insights that can be drawn from the above results. First, there are strong interconnections amongst social and technical (and physical) security factors, attributes, and sub-attributes relevant for modelling and simulating security within critical infrastructure systems and/or organisations. Second, for most security states, there is a combination of one or more technical (and physical) and/or social attributes involved. Quite easily, it can be noted that *a security mechanism is largely an entity either technical or non-technical in form that enforces a security policy to achieve a specific security goal; drawing from the prescribed or desired culture, values, or structure, and being acted upon by an actor or agent*. The enforcement process typically involves the connection or interaction of two or more parts. Third, both social and technical (and physical) attributes interact in a system-like approach, and influence both security and functionality or performance. These attributes co-function to establish and maintain security, and the quality of security reached or achieved depends on the forms and extents to which both social and technical attributes are considered and accounted for in the initial security risk management thinking, design, and building. Fourth, the observed dominant attention on technology advancements even in the security space, and the weak, or absence of involvement of social scientists in the activity of rethinking, reshaping, and re-enforcing information/cybersecurity suggests and explains the common disregard or suppression of social factors and attributes even where they exist in a security model. These are hardly cited and are often in isolation when eventually mentioned.

Clearly, factors and attributes that can cause security weaknesses or strengths within a CNI ecosystem are not solely technical. Social factors and attributes also play key roles. However, the level of attention given to the different factors and attributes can vary, which may suggest the level of perceived significance given to each factor/attribute towards reinforcing or depleting the desired security posture.

The greatest attention and effort on socio-technical security appears to be focusing on the *relationship* or *interactions* between *actors/agents* and *technology hardware and/or software*. It would suggest that security analysts and designers considering socio-technical security approaches appear more concerned with the security issues/risks emerging from actors/agents – humans – traits. A significant proportion of these traits more easily apply to the human actors/agents, than programs, and can relate to that which the actors/agents can or cannot do, that which they are doing, that which they know or have, how they behave, that which drives them, and that which they aim to achieve. Possibly, this could be partly because humans are considered the weakest link in a typical socio-technical interaction system. Another possible reason could link to the increasing cyber incidences that tend to support the weakest-link assertion, based on compromises that have with considerable success, targeted, and exploited the human vector pathway. A third reason may be linked to the ease with which computing systems can be compromised through exploiting the human operators or users, and the damaging consequences and impacts involved.

Past incident reports suggest that human involvements and their direct interactions with technologies within both information and/or operational systems seemingly account for the larger proportion of emerging successful cyber incidents. Thus, newer security modelling and simulation approaches – techniques and tools – need to begin to focus more on how to better understand and represent those human-technology interactions, including the attributes that introduce vulnerabilities in those interactions. It would be helpful to understand the impact of factors such as human security knowledge, security skills, negligence, inappropriate actions or inactions, cognitive overloads, etc., in the light of potential security vulnerabilities and threats that emerge. Appropriate security defence measures and countermeasures also need to be considered, analysed, and modelled. It is easier to reason that modelling security policy attributes would follow next in line after exploring common socio-technical security methods. Again, this would involve translating desirable theoretical security guidelines into practicable solutions.

In terms of security contexts or aspects covered in socio-technical security M&S approaches, results show that a majority of the security modelling works incorporating both social and technical system attributes are focusing on security requirement analysis/engineering for systems or organisations where one or more of human actors/agents, technologies, and organisational elements are involved. Another leading security context that includes modelling and simulation of the interactions between system technology components, concerns tactics for reaching and attacking the components successfully, as well as the attributes of attackers. A third area of focus is the modelling and simulation of threats to the security and privacy of communications data or information.

The works that focus on security requirement engineering such as those by [22], [52], [54], typically cited the use of modelling to reason about, and/or understand the security features needed to secure the various interacting components of an organisation. Undoubtedly, cybersecurity reasoning and solutions need to span people, technology and process elements that co-interact to ensure the normal functioning of a system and/or an organisation in the digital age. This approach supports the much-recommended practice of including and building in security at the conceptual and early design stages of system development. It addresses the typical problems encountered in traditional software development life cycle (SDLC) and traditional cyber-physical/operation technology system developments where security is often considered after systems have already been built and put into operation [75]. This suggests that often, the adoption of cybersecurity in the operational technology (OT) sectors – transport, water, chemical, energy, food, etc., – is driven and motivated by the occurrence of cyber compromises/incidents with the potential for damaging direct and indirect impacts to business functions or loss of life. So, security deployment and/or retrofitting becomes a way to respond to any debilitating outcomes for victim organisations, or a way of guarding against similar damaging impacts for peer organisations not yet affected, who are making efforts towards caution and preparedness. Some of the consequences of this approach to security include high cost of developing and operating such systems, and extended development time [76], [77].

Typically, a key component of security requirement engineering is 'security requirement analysis' or simply 'security analysis' [78]. This describes a process of examining the elements and processes that interact within an organisation or system operations to understand the key sensitivities, functional, operational, and business consequences of security risks (vulnerabilities and threats). Security analysis in OT environments and its outcome are built around availability, integrity, confidentiality, and other associated secondary security principles. Based on these and an understanding of a system/organisation's operational and business objectives, effective safeguards and controls can be identified, selected, and incorporated into system design and development. In addition to enabling a more cost-effective system development process [76], this presents a less stressful, less time-consuming, and more encompassing way to developing, achieving, and maintaining a more secure organisational system and gaining improved security assurance [79], [80]. Using a socio-technical approach ensures that a security requirement analysis is performed with social and technical security risk contexts in mind. It means security views and implications are formed holistically, extending beyond hardware, software, and processes. Security measures and controls are selected and incorporated to capture one or more social contexts amongst actor/agent behaviours, knowledge and skills, interactions, organisational structure, culture, and policies.

The focus on security requirement engineering suggests that the broader and overarching landscape of socio-technical security risks and their associated implications to normal operations do not yet appear to be well-understood by those exploring this space. It highlights initial interests and efforts by security researchers and developers to understand the security needs associated with socio-technical digital systems, the security issues that emerge from interacting socio-technical elements, and how to address the issues prior to practical or actual system development and operations. The motive is assumed to relate to an effort towards getting secure development 'right' from the outset (at design-time) as opposed to the harmful consequences and costs of getting it 'wrong' and having to respond with retrofitting security.

The absence of updated works covering secure system deployment is also an indicator that works in this area are still at their conceptual or theoretical development stages, with real applications not yet in the public domain. This resonates with the results from analysing the extent of application of the socio-technical models reviewed and supports the view that most of the works in this area are still at their 'proof-of-concept' stages. Although not entirely new, socio-technical security modelling and simulation is starting to gain wider traction in research and explorative interests due to evolving security threats. Researchers are still working to understand the dynamics of security interactions and dependencies amongst the range of social and technical factors and attributes identified.

The significant focus on cyber-attack and attacker attributes, strategies, and resource access highlights one of the directions where greater concerns appear to exist and where research seems to be converging. Results reveal a growing desire to understand the attributes of potential cyber attackers, how far into system resources or assets attackers can reach, and the effectiveness of certain attack modes and strategies adopted by cyber attackers. Understanding these contexts can help researchers to identify the subtle malicious schemes of attackers, and to adopt the appropriate security control measures. Thus, the types of security measure(s) adopted are often dependent on the characteristics of evaluated attacks or known attacker capabilities to accomplish specific types of attack on certain system components. Technically, this is referred to as 'threat modelling', and is typically used to identify and prioritise potential security threats, mitigations, and controls to protect system components from sabotage or compromise. Technical security experts – analysts, architects, developers, and managers – use the approach to identify and quantify the gravity of security risk and prioritise appropriate mitigation techniques. Arguably, people with such expertise often come from the computer science and engineering domains. These seem to be more commonly involved in the socio-technical security modelling and simulation works reviewed.

In assessing the maturity of socio-technical security approaches, it is crucial to recognise that there are uncertainties in accurately determining the maturity and accessibility status of some of the socio-technical modelling languages, methodologies, and tools given that they are mostly developed within research domains and used in-house. Reports and documentation on their use and effectiveness are mostly from the same authors, and there is little public evidence of the progressive development and validation, and wider use of these socio-technical modelling approaches in real societal infrastructure. So, whether these socio-technical modelling tools have been discarded, modified, upgraded, or being widely adopted, and at what point; remains information that is not easily accessible in the public domain. Thus, care needs to be taken when considering and drawing inferences and conclusions regarding the maturity of socio-technical security modelling approaches (methodologies and tools) given the possibility of incomplete and imprecise data.

## 4.7   Interdependency Modelling Coverage

In terms of whether dependency attributes are considered in the reviewed works on socio-technical security modelling, it appears dependency attributes are not widely incorporated into current approaches. This may be due to potential difficulties in fully understanding the interaction complexities between the social and technical elements of modern infrastructure systems and their associated security risks. Judging by the various works that explored this from a solely technical or social point, reasoning and modelling component dependencies may seem easier to accomplish. For example, there are a range of works that have explored the quantitative modelling of complex infrastructure systems cascades [81], [82].

From a socio-technical point of view, some security modelling approaches incorporating dependency modelling appear to be emerging too. Some common examples from this study include; The CySeMoL Socio-Technical Modelling Language and Tool [51] designed and used for enterprise-level system architecture modelling and vulnerability assessment. Built on a prior framework [83], the tool uses a probabilistic relational model (PRM) [84] to support security analysis. The CySeMoL's PRM consists of both logical deterministic (first order and second order) and probabilistic dependencies with uncertain influence. These dependencies can be used to support the estimation of the probability that a penetration testing professional using publicly available tools, can successfully attack and compromise a prescribed architecture within a set period, e.g., one week.

Another socio-technical security modelling approach found is 'Secure Tropos' which is an extension of the *'Tropos'* methodology and which includes security features based on the concept of security constraints [52]. This implies that the *'Tropos'* concepts of dependency, goal, task, resource, and capability were extended with security in mind and formed the secure entities of *'Secure Tropos'*. This has enabled several security modelling functions including security reference modelling, security constraint modelling, secure entities modelling, and secure capability modelling [52].

Another approach that considered dependency modelling is the SePTA (Security, Privacy and Trust Approach) method which is said to support a unified specification of security, privacy, and trust requirements under one framework, and enables security experts to implement such requirements. SePTA is said to be more suitable for complex socio-technical information systems such as the types used in public and large institutions, where collaborative interactions exist between people and autonomous technical components. Although interaction complexities in the social and technical system ensemble can pose serious complications, some understanding of the dependencies amongst elements in these two categories can enable better identification of security risks and the design of more effective security response strategies and controls [85]. Also, security dependency considerations and M&S can provide a capacity to understand and prepare against the potential harmful consequences and cascading impacts that can occur as a result of security compromises in relation to system cross-dependencies.

# 5. Conclusion

## 5.1 Summary

Critical National Infrastructure (CNI) systems are fundamental, and critical in providing and maintaining vital societal functions, including health, safety, security, economic, and social well-being of people. CNIs are critical because disrupting or destroying them or their functions would have serious physical, economic, and social consequences. To guard against these types of occurrences, protective measures are required. There is significant progress in the advancement of security solutions to address the increasing occurrence of cyber-attacks and successful mal-interventions, especially those targeting CNI systems. Technologies in fields of information, communications, and networking continue to advance, and cyber-malicious activities and incidences are in lock-step with these. However, it is clear that technology alone is neither the source nor the solution for many of the emerging cybersecurity challenges impacting CNI systems. Thus, the heavy focus on technical security solutions alone or in isolation by the security community is holding them back in enabling more efficient response capacities to address the evolving activities of modern malicious cyber threat agents. At minimum, CNI systems need digital/cybersecurity jointly covering technology infrastructure (tangible and intangible), people/personnel, data/information, and organisational structures, all at the same time.

Considering cybersecurity within CNI sectors such as Transport, Water, Energy, Nuclear, Defence, Chemicals, Healthcare, and Food & Agriculture really means thinking about protecting the *'real world of dangerously-impacting technologies'*. It is a world of rigorous security design and development, with a foremost objective of *'get it right first time'*. The cost of getting it wrong can be huge, nearly similar to the cost of making fundamental changes and updates. It is not desirable to find security issues and risks shutting down traffic signals on roads and train tracks, reconfiguring or causing mal-operation of controllers and actuators driving chemical mixing or supporting manufacture and/or distribution of fuels, water, wastewater, even worse, shutting down communication networks and safety systems in process plants, etc. The consequences of these can be huge and adversely overwhelming. So, while it may not be possible to completely eliminate these risks directly in the real environment, they can be mitigated by exploring defences based on a well-informed understanding of the effectiveness of available mitigations and their implications. In addition, this can inform an understanding of how cyber resilience of those CNIs can be achieved in line with the national cybersecurity strategy.  But the existing IT and network communications processes around security audit, testing, and evaluation are too risky to be performed on real-world CNI systems. This is because things can go wrong that are capable of threatening human, property, societal and/or environmental safety, and/or revenues and reputation. Modelling and simulation (M&S) provide the alternative approach to achieve similar levels of security auditing and testing, thus gaining assurance with a greatly reduced possibility of directly interrupting critical functions or operations.

The Blackett Review [86], *"modelling is going through a revolution"* was partly driven by the dramatic increase in the availability of computing power including cloud innovations. The modelling revolution is driven by deeper, and better reasoning and understanding about the composition of CNIs as not solely comprising technologies but viewed as socio-technical systems; their interactions and individual contributions forming a functional ecosystem. Socio-technical CNI comprises a co-integration of engineering technologies and systems (hardware, software, and processes) that require interactions with human actors/agents and some social and organisational attributes (governance, policy, regulations, etc.) to perform their function or achieve a desired operational objective. Thus, understanding the functionality, reliability, strengths, and weaknesses of such systems relies on an analysis that spans the technological and social domains. The security risk attributes – threats, vulnerabilities, impacts – of modern CNI systems also bear technical and social dimensions, and technical solutions alone are not sufficient to support and assure the best protection and resilience required. Human factors such as security awareness and training (knowledge and skills), attitudes, cognitive overloads, and behaviours, as well as organisational factors such as governance, policy, and regulation, leadership, communications, culture, etc., are also crucial and complementary. Thus, a socio-technical approach to security reasoning and design via M&S can support a more effective outcome for CNI security and resilience. Although infrastructure interdependencies can mean that this may seem more complicated to achieve than focusing on singular infrastructure and the technical or social elements separately, a combined approach is worth trying in order to understand the extent to which security can be modelled socio-technically, and the extent to which positive security impact is possible.

Most applications of M&S in critical infrastructure security appear more focused on learning about and addressing technology-based security issues. This is changing however, with literature and several works emerging that look into socio-technical security and exploring how this might be achieved using M&S. This is an indication that the cybersecurity community is awakening to the reality of the unfolding and evolving nature of security risks, and are not only concerned, but also thinking, talking, and exploring potential solutions. Clearly, to address the issues of evolving cybersecurity risks and the widening cyber threat landscape impacting critical infrastructure sectors, a rethink of the approach to security analysis, modelling, simulation, and implementation is required.

A cybersecurity approach that considers and includes broader cybersecurity risk factors by combining both social and technical (and potentially environmental) security attributes and countermeasures would be beneficial. Early analysis of potential security problems from socio-technical viewpoints will be more beneficial for modern CNI system design, which can result in the development of systems that have better robustness. We refer to this as a *'Socio-Technical Security by Design (STSbD)'*, used to describe the security reasoning, modelling, and implementation approach that involves considering and combining technology factors (hardware, software, processes, etc.,), with human factors (actors, actions or inactions, knowledge, skills, behaviour, etc.,), and organisational aspects (structure, culture, policies, etc.,). This can help to achieve a deeper understanding of the interactions and interdependencies amongst these elements and attributes in the socio-technical CNI ecosystem, and which can make or mar security. An *'STSbD'* approach can help to reach a broader analytical coverage of CNI interacting system elements, factors, and attributes that influence security risks, and support the creation of more effective security countermeasures or solutions, beyond a solely technical view. Security aspects of CNI design need to encapsulate the broader picture including all the co-interacting elements that enable critical services.

Depending on scope, where possible, the needs and actions of the various socio-technical CNI system constituent elements participating in the complex interactions should be considered and captured to help address the unique security concerns and roles within such dynamic environments. For security M&S in CNIs, it is necessary to understand how to achieve reflective security goals that combine technical (plus physical) and social attributes. This can lead towards a more holistic security model, enabling the evolution of higher-level national and international guidance on future developments and improvements in the implementation of modern, IoT-enabled CNI security systems. For UK critical infrastructure organisations, this includes checks to ensure compliance with core government security requirements, best practices, and standards that reference the Governments' *Minimum Cybersecurity Standards* [87], the NIS directive-led NCSC Cyber Assessment Framework (CAF) [88], and *Technology Code of Practice* (TCOP) [89], which cover security for people, process, and technology.

Thus, those who own, operate, and manage CNI systems need to understand their unique socio-technical environments, so that they can become more aware of emerging cybersecurity threats, vulnerabilities, risks, and harmful impacts, as well as the exploratory options available to help them enhance cybersecurity, resilience, and safety. Awareness needs to be improved in the CNI community regarding state-of-the-art socio-technical security modelling and simulation approaches, and how these can enable better reasoning about, and mitigation of security weaknesses, especially when IoT and other emerging technologies are being applied.

Government and policymakers need to understand how to support the development of new and improved socio-technical security M&S technologies and methodologies, and how to promote their practical use within CNI sectors. They need to understand policy interventions that can further contribute to shaping and signposting the cybersecurity of CNIs, and how decision-making on more effective M&S approaches – tools and techniques – can improve security in modern CNI systems. Further to understanding, policy stakeholders need to also take practical steps to support the use of security M&S approaches to help bolster the cybersecurity and resilience capacity of CNI owners and operators against cyber-attacks. This can include raising awareness as highlighted initially, tailored security training at all levels, and also sharing best practice guidelines. This re-emphasises a key recommendation from the *'Data for the Public Good' Report* [90] by the National Infrastructure Commission in 2017, leading to the role of the National Digital Twin programme (NDTp) in simplifying the management (including the security) of physical infrastructure.

Cybersecurity and resilience in CNI systems need to be socio-technical by design. To get the best out of a socio-technical approach for M&S, users must work closely with security modelling analysts and developers throughout system model creation and application. This is an essential factor in establishing confidence on what a socio-technical security model can and cannot do – its maturity, usability, and effectiveness.

Considering the state-of-the-art, application, and maturity of socio-technical security modelling approaches in CNI systems, demonstrating the application or use of a security model in a test or real system/environment can provide a stronger and more convincing case or proof of maturity. A security model is meant to address a security problem in a real/live environment. Thus, there is hardly a better way to demonstrate initial maturity of a security model except by implementing the model in a realistic test or actual environment and showing that it works to establish what is claimed or desired. This would imply that the security model in question has undergone rigorous and perhaps multiple reviews and analyses of functionalities and failures, that weak points have been identified and fixed following tests activities, and possible improvements made, to achieve a more robust scheme or model. Amongst the socio-technical security modelling approaches reviewed, the *'CySeMoL'* tool appears to reflect these characteristics and can be considered more mature than others reviewed.

However, given that the attacks and countermeasures covered in the *'CySeMoL'* use case leading to security M&S outputs are originally drawn from literature and inputs from security domain experts, the quality of the outcome may depend on the level of input expertise. Thus, the wider capability and effectiveness of the tool will depend on continuous updates or version upgrades following qualitative inputs from literature studies, experiments, and structured expert elicitations. These will then influence the derivation of parameters for deterministic and probabilistic dependences. There are also uncertainty issues with the correct representation of security viewpoints as qualitative evaluations and feedback that imply a level of subjectivity. This can raise questions around the consistency and reliability of M&S outcomes, and the extent to which the outcomes can be accepted and used to inform decision-making.

Often, security models evolve towards maturity following numerous acknowledgements and critiques, which inform further improvements that address known gaps and new functionalities. From the reviewed socio-technical security approaches, the *'STS-ml'* approach and other socio-technical security modelling approaches appear to be moving towards maturity for this reason. Acceptance can be further strengthened with evidence of contextual validations using at least realistic test environments to show that these socio-technical security modelling approaches are usable, and support claimed security objectives. This helps to build credibility for the security M&S tool and its outputs in use. Knowing that a socio-technical security model has been applied to a real system or environment and that the results are promising, can reduce the fears or concerns related to the consideration or possibility of things going wrong, or of being misguided by outcomes of using the security tool.

While some common security assessment standards such as the ISO/IEC 27000:2018 [91] and the security metric guide from NIST [92] emphasise the requirement for users to describe how organisations can build and drive security assessments, the varied security assessments and their interpretations are still unclear. Thus, their effectiveness still depends on user expertise. This can be a problem for low-level technical security users in a socio-technical security modelling, simulation, and analysis context. That is why, from a usability point of view, a socio-technical security M&S approach (tools, techniques, or methodology) need to be easy to use, and not require high-level technical security expertise from users. Security M&S approaches along these lines should be able to assist users to easily achieve a finished aggregation structure or scheme for security risk or countermeasure properties. Some of the reviewed socio-technical security modelling approaches such as the 'Secure Tropos' method, and others that use tree-structures (for attack and defence analysis) appear to be adopting the usability characteristics described above. This helps users to combine multiple security variables or principles into a single overarching security goal - e.g., determining total vulnerabilities or impacts and producing a detailed and reliable report - without needing to know or perform the individual technical analysis themselves.

Regardless, efforts need to be furthered towards reducing the requirements for security expertise while allowing for security evaluations. For example, the authors of the *CySeMoL* tool claim that deep security expertise is not required by users. Users of the security M&S approach and tool only need to supply information about the system's architecture and do not need to provide further security-related attributes such as attack probability or mean-time-to-compromise values. Rather, system security properties are derived from the system architecture information provided and computed according to a common evaluation theory. This means that system designers and developers have a socio-technical security M&S and analysis tool available and can benefit from understanding the basic security needs and requirements to consider and incorporate into critical infrastructure system-building. The security M&S tools along this usability pathway encourage consideration of the multi-disciplinary nature of contexts around socio-technical security thinking and would support greater adoption by non-technical security experts.

## 5.2   Areas of Further Research

For further research, it would be interesting and useful to explore the effectiveness of existing methodologies like the *'Secure Tropos'* for reasoning about, and modelling socio-technical security requirements in critical infrastructure software development process. Furthermore, exploring or demonstrating the methodology or tool's application and impact in real critical infrastructure security domains can enhance credibility and acceptance by the security modelling and simulation community.

The claims that the *'CySeMoL'* socio-technical security modelling language performs well in modelling enterprise IT-level system architecture and vulnerability analysis also need to be further researched and/or validated. Potential extensions of the tool to explore modelling capabilities for industrial control (OT level) system architectures will be a significantly beneficial improvement. Also, applying the *'CySeMoL'* tool to explore security use cases in other critical infrastructure sectors such as rail transportation, water and wastewater distribution systems, and energy, will be highly beneficial to learn its effectiveness across the CNI domains.

A more encompassing capability for socio-technical security M&S may also be realised by exploring tools, techniques, and methodologies that combine multiple analysis viewpoints. An example is seen in the *SePTA* methodology that is said to combine four modelling languages/methods (STS-ml) [22], [25], [66], SecBPMN2 [93], Secure Tropos (SecTro) [94] and JTrust [95] by applying transformations that enable the individual tools to exchange models. Individual tools contribute to the overall security requirement engineering process and objectives by performing a phase or subcategory of security analysis.

Another hypothetical example suggested involves merging the capacities of *'Portunes'*, *'Secure Tropos'*, and the *'Actor-Network Theory'*. This offers the potential to achieve something better than each of them separately. A combination might yield an outcome where the three tools can work together as one with a socio-technical security modelling capacity that is greater than each one in isolation or the sum of the parts. The ensemble would have three most basic capacities. First - enabling the modelling of the wider CNI system components including physical, digital, and social attributes, their interactions, and potential attack approaches or modes. Second - supporting a socio-technical security requirement analysis to drive CNI system/component design and implementation, and third - adding an ability to support reasoning and representation of human actions in the CNI security access management contexts. The latter is important because the effectiveness of attacks or defences often depend on the unpredictable response/actions of humans-in-the-loop of CNI systems.

It must be noted that it is not ultimately possible to have bullet-proof cybersecurity, that enables to completely avoid or prevent cyber-attacks and/or intrusions to modern CNIs with functionally or operational disruptive or destructive consequences. Thus, cyber resilience should aim for a balancing objective of compensating when or where cybersecurity fails. The ability to recover and remain operational in the face of a cyber incident is increasingly crucial. This re-emphasises why part of the updated UK Government's new Cybersecurity Strategy 2022-2030 [96], highlights a vision to ensure that core government functions – from public services delivery to the operation of National Security systems – are resilient to cyber-attacks. There is an understanding that it is not a matter of if, but when cyber-attacks will happen, hence the need to prepare by building in greater cyber resilience to harden critical functions. This requires the use of M&S to explore and assure the activities, emphasising the five objectives that define the dimensions of cyber resilience targeted. Noting that the risks targeting CNIs including government assets are not solely technical, it would be good to explore how socio-technical security analysis and M&S can be used to help realise UK government's cyber resilience objectives of managing cybersecurity risks, protecting against cyber-attacks, detecting cybersecurity events, building cybersecurity awareness, knowledge, skills, and culture, and minimising the impact of cybersecurity incidents.

## 5.3  Recommendations

Given the above views and insights, it is crucial to apply an evolutionary approach involving model system design creation, testing, and refinement, and that it be adopted and followed to achieve and maintain a more holistic, robust, and resilient security capacity within industrial infrastructure systems. This is because achieving (through design and implementation) suitably secure and resilient socio-technical critical industrial/cyber-physical infrastructure control systems cannot solely happen at requirements analysis or at design stages. Nor can it be achieved from a one-time implementation of a system. Quite often, in a structured industrial environment, social and technical security factors tend to mutually influence each other, thus it is impossible to analyse and address all the potential effects of technology on a social system and vice versa. An evolutionary approach socio-technical security reasoning, design, and implementation allows for continuous evaluations and joint-optimisation of security and resilience in response to the dynamics of security risks.

To realise the goal of robust socio-technical security (and resilience) during CNI system-thinking and development, there should be improved communication amongst stakeholders about socio-technical security issues and risks. This is crucial to enable constructive engagement/support for considering and using the information about socio-technical security factors and attributes in CNI system designs and processes for organisational changes.

Communication can take the form of awareness and briefing activities to CNI stakeholders relating to the functionality and security concerns of their colleagues across the CNI operational system. This pertains to both maintaining and disenabling security and offering compelling explanations on the usefulness of adopting a socio-technical security-thinking and M&S approach, prior to live system deployments. For example, in this time of technological advancement and IT-OT convergence, operational technology engineers involved in the design of safety systems might need to be informed of the impossibility of achieving robust safety implementations without considering security attributes. Also, enterprise/IT systems engineers/managers responsible for designing/using databases might need to understand that not all types of data are feasibly accessible and collectible in the OT environment. Similarly, security designers and architects need to be aware that designing in very strong security measures can make OT systems impracticable or difficult to use by operational-level engineers and technicians, which can further inhibit performance and productivity.

Our prior research on improving the security of transport infrastructure efficiency systems showed that progressing this type of communications amongst CNI system stakeholders can help bridge the gap in socio-technical system development and change processes. It helps to feed information on socio-technical security issues to CNI system development teams and enables a capability to use such information in driving timely and constructive decision-making on more effective security design solutions.

Better security can be achieved by understanding CNI system and component functions and interactions in the context of implementable security mechanisms. This means decomposing functional and state requirements to identify relevant security features and structuring the features in a form that can be easily translated into active security mechanisms within the system. This can be a good way to embed security into software and/or hardware used in digital or cyber systems to thus enacting security-by-design. Such decomposition and analysis also help ensures that security considerations around potential vulnerabilities, threats, and impacts from the human-technology-physical interactions and expectations are properly considered from the outset of system development. It also ensures that the necessary controls and plug-ins necessary to establish the desired security expectations are included from the outset. By these means, typical integration issues around component compatibilities and functionality disruptions often related to cyber-physical systems can be more easily addressed or eliminated.

Government can help ensure the wider adoption of socio-technical security by developing policies that can shape and signpost the CNI cybersecurity environment, including how decision-making concerning socio-technical analysis and M&S approaches (techniques and tools) can enhance security in modern critical infrastructure systems. Better outcomes can be achieved if policies concerned with signing off projects include the provision of evidence that security around socio-technical-physical systems interactions are well captured at design stages before implementation starts. Showing that broader socio-technical security factors and conditions have been considered and captured in security requirements analysis and design can strengthen confidence and trust in the holistic nature of the system to support the decisions and approval for implementation.

# Appendix A: Reviewed Literatures on Socio-Technical Security Modelling and Simulation Approaches

| Authors | Authors Affiliations | Article Title |
| --- | --- | --- |
| Dragovic and Crowcroft 2004 | University of Cambridge | Information Exposure Control through Data Manipulation for Ubiquitous Computing |
| Dragovic and Crowcroft 2005 | University of Cambridge | Containment: From context awareness to contextual effects awareness |
| Dragovic 2006 | University of Cambridge | CASPEr: Containment-Aware Security for Pervasive Computing Environments |
| Mouratidis and Giorgini 2007 | University of East London & University of Trento, Italy | Secure Tropos: A Security-Oriented Extension of the Tropos methodology |
| Probst et al 2007 | Technical University of Denmark | Where Can an Insider Attack? |
| Probst and Hansen 2008 | Technical University of Denmark | An extensible analysable system model |
| Mathew et al 2008 | State University of New York at Buffalo | Insider abuse comprehension through capability acquisition graphs |
| Franqueira et al 2009 | University of Twente | Multi-step attack modelling and simulation (MsAMS) framework based on mobile ambients |
| Probst and Hansen 2009 | Technical University of Denmark | Analysing Access Control Specifications |
| Dimkov et al 2010 | University of Twente | Portunes: Representing Attack Scenarios Spanning through the Physical, Digital and Social Domain |
| Pieters 2011 | University of Twente | Representing Humans in System Security Models: Representing Humans in System Security Models: An Actor-Network Approach |
| Pavlidi and Islam 2011 | University of East London | SecTro: A CASE tool for modelling security in requirements engineering using Secure Tropos |
| Dalpiaz et al 2011 | University of Trento | Security Requirements Engineering via Commitments |
| Sameer 2011 | Technical University of Denmark | Attack Generation from System Models |
| Kowalski and Mwakalinga, 2011 | Stockholm University | Modelling the Enemies of an IT Security Systems - A Socio-Technical System Security Model approach |
| Sommestad et al 2013 | Royal Institute of Technology, Stockholm | The Cyber Security Modeling Language: A Tool for Assessing the Vulnerability of Enterprise System Architectures |
| Paja et al 2013 | University of Trento & University of Toronto, Canada | Specifying and Reasoning over Socio-Technical Security Requirements with STS-Tool |
| Paja et al 2013b | University of Trento & University of Toronto, Canada | Designing Secure Socio-Technical Systems with STS-ml |
| Paja et al 2014 | University of Trento | STS-Tool: Security Requirements Engineering for Socio-Technical Systems |
| Lenzini et al, 2015 | University of Luxembourg | Security analysis of socio-technical physical systems **(LM015)** |
| Ionita et al 2015 | University of Twente | Investigating the usability and utility of tangible modelling of socio-technical architectures |

| Authors | Authors Affiliations | Article Title |
|---|---|---|
| Aslanyan et al 2015 | Technical University of Denmark | Modelling and Analysing Socio-Technical Systems |
| Ivanova et al 2016 | Technical University of Denmark, Aalborg University & Middlesex University, London, UK | Transforming Graphical System Models to Graphical Attack Models |
| Ostwald 2017 | Salient Works (Industry), Kirkland, Washington | Threat Modeling Data Analysis in Socio-technical Systems |
| Mujinga et al 2017 | University of South Africa | A Socio-Technical Approach to Information Security |
| Zhou et al 2018 | Liverpool John Moores University & Zhejiang University, China | A 3D Security Modelling Platform for Social IoT Environments |
| Zoto et al 2018 | Norwegian University for Science and Technology | Using a socio-technical systems approach to design and support systems thinking in cyber security education |
| Salnitri et al 2020 | Politecnico di Milano, Milan & Collibra (Industry) Brussels-Belgium, & University of Brighton, UK, & University of the Aegean, Samos-Greece, University of Trento, Italy | Modelling the interplay of security, privacy and trust in sociotechnical systems: a computer-aided design approach |